\documentclass[a4paper,11pt]{article}
\pdfoutput=1
\usepackage{jcappub}

\usepackage{amsmath,amssymb,graphicx}
\usepackage{soul}
\usepackage[latin1]{inputenc}
\usepackage{dsfont}

\usepackage{subfigure}

\usepackage{array}
\usepackage{mathtools}
\usepackage{bbold}
\usepackage{float}

\usepackage[thinlines]{easytable}

\newcommand{\be}{\begin{eqnarray}}
\newcommand{\en}{\end{eqnarray}}

\newcommand{\Hubble}{ H }

\usepackage[section, below, above]{placeins}

\usepackage{afterpage}

\begin{document}

\begin{titlepage}

\vspace*{-15mm}
\vspace*{0.7cm}

\begin{center}

{\Large {\bf Non-thermal Leptogenesis after Majoron Hilltop Inflation}}\\[8mm]

Stefan Antusch$^{\star\dagger}$\footnote{Email: \texttt{stefan.antusch@unibas.ch}},  
Kenneth Marschall $^{\star}$\footnote{Email: \texttt{kenneth.marschall@unibas.ch}},

\end{center}

\vspace*{0.20cm}

\centerline{$^{\star}$ \it
Department of Physics, University of Basel,}
\centerline{\it
Klingelbergstr.\ 82, CH-4056 Basel, Switzerland}

\vspace*{0.4cm}

\centerline{$^{\dagger}$ \it
Max-Planck-Institut f\"ur Physik (Werner-Heisenberg-Institut),}
\centerline{\it
F\"ohringer Ring 6, D-80805 M\"unchen, Germany}

\vspace*{1.2cm}

\begin{abstract}

\noindent We analyse non-thermal leptogenesis after models of Majoron hilltop inflation, where the scalar field that provides masses for the right-handed neutrinos and sneutrinos via its vacuum expectation value acts as the inflaton. We discuss different realisations of Majoron inflation models with different hilltop shapes and couplings to the right-handed (s)neutrinos. To study the non-thermally produced baryon asymmetry in these models, we numerically solve the relevant Boltzmann equations. In contrast to previous studies, we include the effects from resonant sneutrino particle production during preheating. We find that these effects can result in an enhancement of the produced  baryon asymmetry by more than an order of magnitude. This can significantly change the favoured parameter regions of these models.

\end{abstract}
\end{titlepage}

\section{Introduction}

Inflation \cite{inf} is a successful paradigm for the early universe, able to explain the initial conditions of the hot big bang, i.e.\ resolving the flatness and horizon problems. During inflation the universe undergoes a phase of accelerated expansion driven by a scalar field, the so-called inflaton. One challenge for this scenario is that the exponential expansion during inflation would dilute any preexisting baryon asymmetry, which means that a mechanism is required to generate it after inflation. Another challenge for inflation models is the fact that they have to preserve a rather flat inflaton potential for the inflationary phase, but also have to provide a coupling to other fields which allow for reheating into, finally, the Standard Model fields. 

Both challenges can be resolved in models of supersymmetric Majoron hilltop inflation, where the scalar field that provides masses for the right-handed neutrinos and sneutrinos via its vacuum expectation value acts as the inflaton field which drives inflation (see e.g.\ \cite{Senoguz:2004ky,Nakayama:2011ri,Antusch:2014qqa, Boucenna:2014uma,King:2017nbl}).  
We focus on scenarios where the inflaton potential has a ``hilltop-shape'' (as in the realisations of \cite{Senoguz:2004ky,Nakayama:2011ri,Antusch:2014qqa, King:2017nbl}). Its flatness is protected against large radiative corrections due to the underlying supersymmetry. 
The coupling that leads to the right-handed (s)neutrino masses automatically provides the dominant decay mode of the inflaton. The successive decays of the right-handed (s)neutrinos into Higgs(inos) and (s)lepton doublets then allows for efficient reheating and can also generate the baryon asymmetry of the universe via non-thermal leptogenesis \cite{leptogenesis}. 

In this paper, we study the prospects for generating the baryon asymmetry in such models via non-thermal leptogenesis during the reheating process. We systematically discuss realisations of Majoron hilltop inflation models and their reheating phase, including higher power operators of the inflaton field, as well as different types of operators for the generation of the right-handed neutrino masses. We provide results for the generated baryon asymmetry and the reheat temperature and discuss the differences between the various model realisations.
Compared to earlier papers, which discussed non-thermal leptogenesis in specific models (cf.\ \cite{Senoguz:2004ky,Antusch:2014qqa}), we include effects of preheating, in particular the non-perturbative production of sneutrino fluctuations via parametric resonance, using the results of the lattice simulations from \cite{Antusch:2015vna}. 

The paper is organised as follows: In section \ref{S1} we introduce the class of Majoron hilltop inflation and reheating models and review its predictions for the spectral index and the tensor-to-scalar ratio. Then we consider the time after inflation where we discuss effects of preheating, followed by perturbative inflaton decay. In section \ref{S2} we describe the calculation of the baryon asymmetry via the non-thermal leptogenesis mechanism. In section \ref{S3} the results for the produced baryon asymmetry and estimates for the reheat temperature are presented. The Appendix contains details on the decay rates of the inflaton and the right-handed (s)neutrino,

\section{Majoron Hilltop Inflation Models}\label{S1}

\begin{figure}[t] 
\centering
\includegraphics[width=14cm,height=12cm,keepaspectratio]{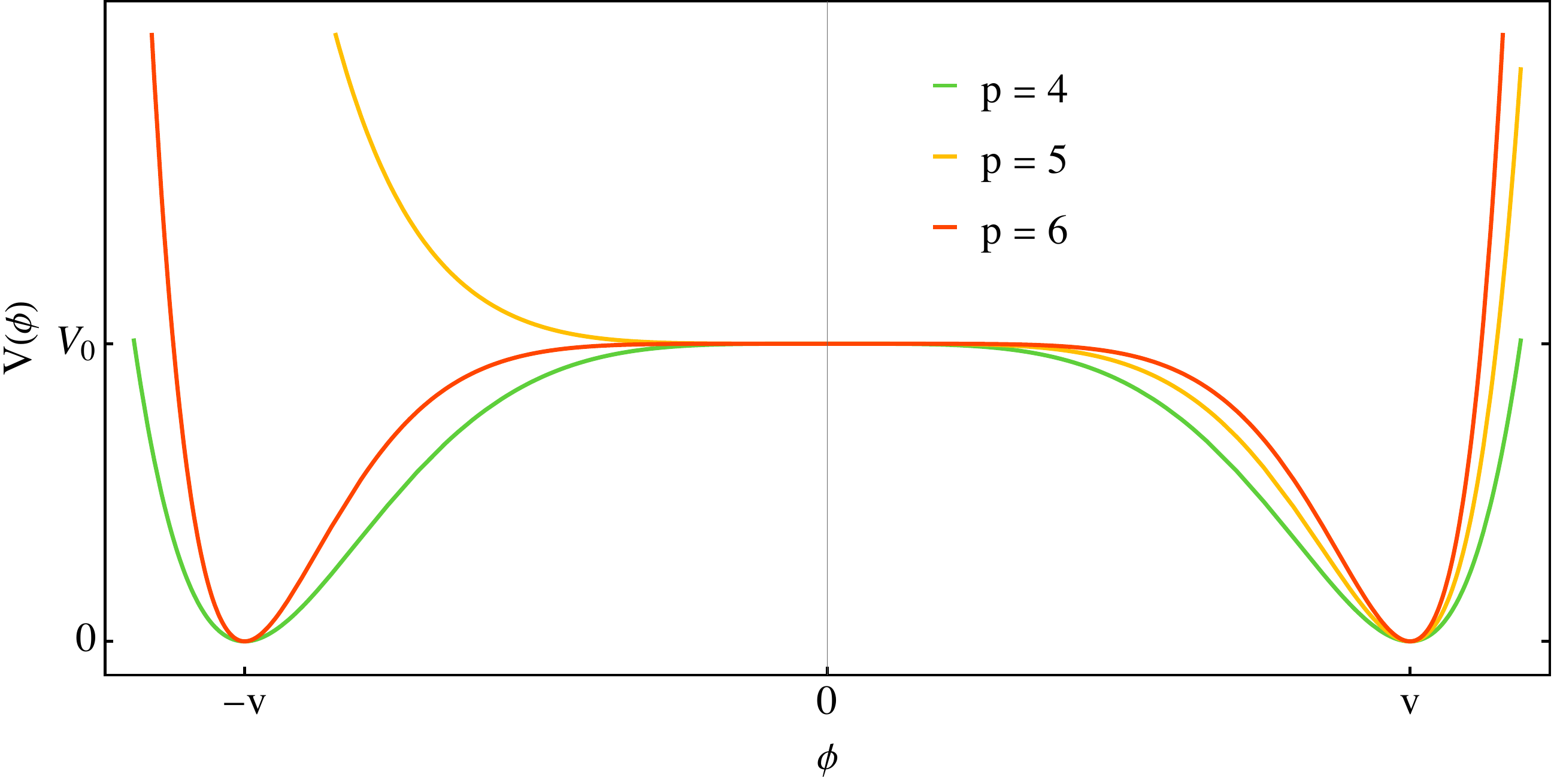} 
\caption{Illustration of the real part $\phi=\sqrt{2}Re[\Phi]$ of the inflaton potential. During inflation the inflaton rolls slowly from close to the top of the hill (where $V(\phi) \approx V_0$) towards its minimum at $\phi = v$.}
\label{fig:1}
\end{figure}

Different realisations of Majoron hilltop inflation models with different hilltop shapes for the inflation phase and different couplings to the right-handed (s)neutrinos for reheating and leptogenesis can be realised in the framework of supersymmetry/supergravity with the following superpotential,
\begin{equation}\label{B1}
W=\sqrt{V_0}\,\hat{S}\,\biggr(1-\frac{2^\frac{p}{2} \,\hat{\Phi}^p}{v^p}\biggr)+\lambda_i\,\hat{\Phi}^n \hat{X}_i^2+y_{ji}\hat{L}_j\hat{H}_u\hat{X}_i \:,
\end{equation}
which extends the superpotential of, e.g.\ the Minimal Supersymmetric Standard Model (MSSM). The chiral superfield $\hat{\Phi}$ contains the inflaton $\phi=\sqrt{2}Re[\Phi]$ as scalar component, where the imaginary part has been set to zero as discussed in \cite{Nolde:2013bha, Antusch:2014qqa}. $\hat{X}_i$, $i=1,\dots,3$ are the right-handed chiral neutrino superfields, with the complex scalar fields $X_i$ and the fermionic right-handed neutrino $\psi_{\chi_i}$ as components. The third term represents the Yukawa couplings to the MSSM Higgs doublet $\hat{H}_u$ and the (s)lepton doublet superfields $\hat{L}_j$. $\hat{S}$ is a ``driving superfield'' for which we will fix its scalar component at $S=0$ (during and after inflation) by a suitable non-canonical term in the K\"ahler potential (for a discussion, see e.g.\ \cite{Antusch:2014qqa, Antusch:2008pn}).  
In the following we assume that only the right-handed (s)neutrino $\hat X_1$ is lighter than the inflaton $\hat{\Phi}$ such that we can ignore $\hat X_2$ and $\hat X_3$ in the analysis. We will refer to $\hat X_1$ simply as $\hat X$ from now on. 
For simplicity we assume a canonical K\"ahler potential for $\hat X$ and $\hat{\Phi}$.\footnote{As discussed in \cite{Antusch:2013eca}, models of this class can also be combined with a K\"ahler potential featuring a Heisenberg symmetry, which can resolve the $\eta$ problem (i.e.\ further protect the flatness of the potential).}

\subsection{Inflation Phase}
\begin{figure}[t!] 
\centering
\includegraphics[width=14cm,height=12cm,keepaspectratio]{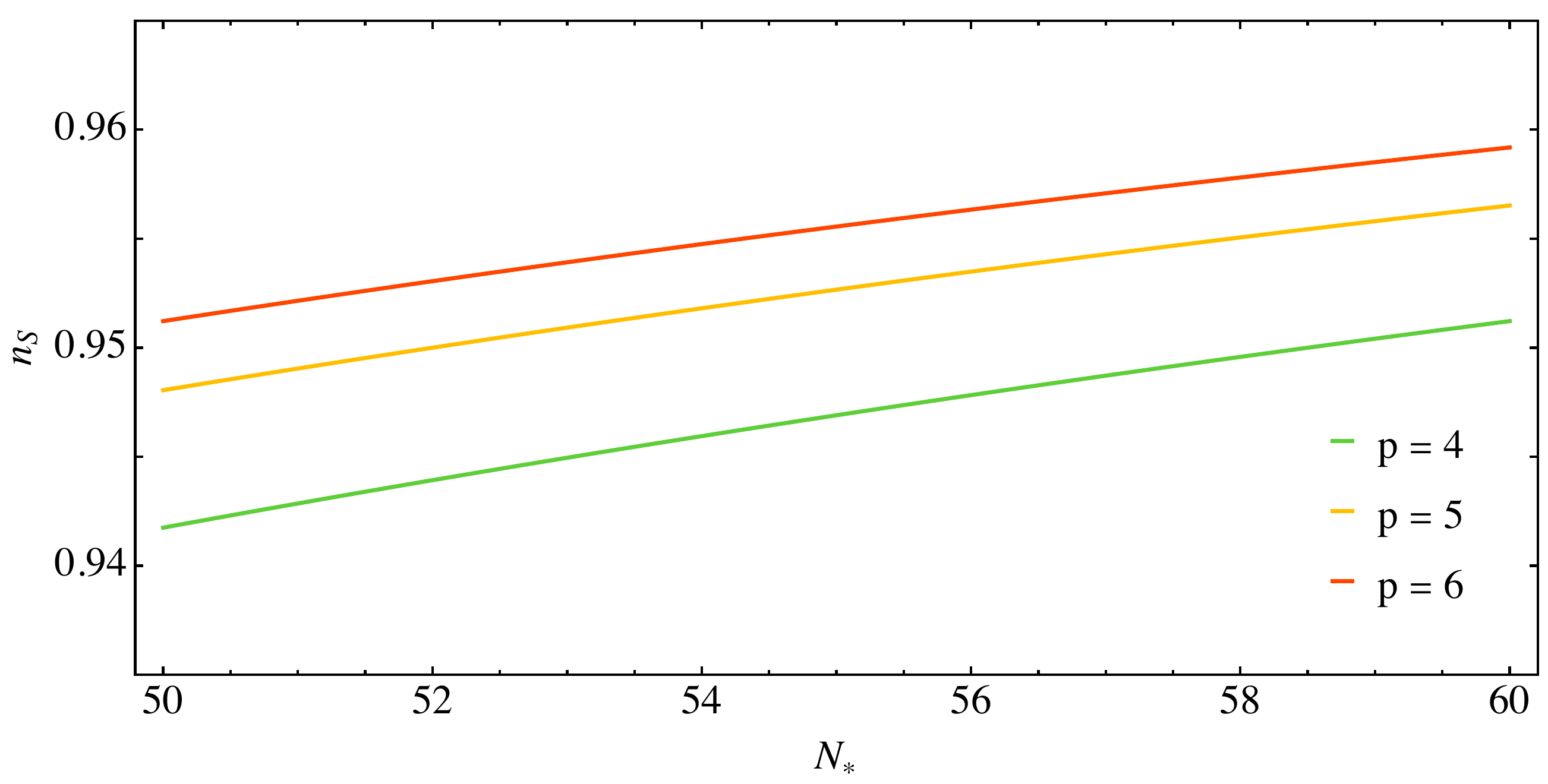} 
\caption{Predictions of the spectral index $n_\mathrm{s}$ in terms of $N_*$ for different values of $p$.}
\label{fig:2}
\end{figure}

From Eq.~(\ref{B1}) and with $S = 0$ we obtain the following scalar potential for $\phi$ and $\chi$ (with $\chi :=\sqrt{2}\, |X|$):
\begin{equation}\label{B2}
\begin{aligned}
V(\phi,\chi)&=\,\biggr\rvert\frac{\partial W}{\partial \hat{S}}\biggr\rvert_{\theta=0}^2+\,\biggr\rvert\frac{\partial W}{\partial \hat{\Phi}}\biggr\rvert_{\theta=0}^2+\,\biggr\rvert\frac{\partial W}{\partial \hat{X}}\biggr\rvert_{\theta=0}^2\\&
=V_0\,\biggr(\,1\,-\,\frac{\phi^p}{v^p}\,\biggr)^2+\,\frac{\lambda^2 n^2}{2^{n+1}}\,\phi^{2n-2}\, \chi^4\,+\,\frac{4\lambda^2}{2^{n+1}}\,\phi^{2n}\, \chi^2+...\: .
\end{aligned}
\end{equation}
During inflation $\chi$ has zero field value and the potential for different values of $p$ has the form shown in figure \ref{fig:1}. Inflation proceeds when the field $\phi$ is close to the top of the ``hill", i.e.\ close to $\phi = 0$, and ends when 
\begin{equation}\label{B3}
\phi_{end}=\biggr(\frac{v^p}{2\,p\,(p-1)m_{pl}^2}\biggr)^\frac{1}{p-2},
\end{equation}
where $\eta \equiv m_{pl}^2(\partial^2 V/\phi^2)/V)\simeq -1$. With $N_*$ denoting the number of $e$-folds between the horizon exit of relevant perturbations and the end of inflation the field value at horizon crossing, $\phi=\phi_*$ can be deduced:  
\begin{equation}\label{B4}
\phi_{*}=\biggr(\frac{2\,p\,m_\mathrm{pl}^2}{v^p}((p-2)N_{*}+(p-1))\biggr)^\frac{1}{2-p}.
\end{equation}
By evaluating the slow-roll parameters $\varepsilon_\phi = \frac{1}{2}m_\mathrm{pl}^2(\partial V/\partial\phi)^2/V^2$ and $\eta_\phi$ at $\phi=\phi_*$, the prediction for the spectral index $n_\mathrm{s}$ and the tensor-to-scalar ratio $r$ are given as:
\begin{eqnarray}\label{I5}
n_\mathrm{s}\,\, &=& 1-6\,\varepsilon_\phi(\phi_{*})+2\,\eta_\phi(\phi_{*})\simeq 1+2\,\eta_\phi(\phi_{*}),\\[0.4cm]
r\,\, &=& 16\,\varepsilon_\phi(\phi_{*}),
\end{eqnarray}
where
\begin{eqnarray}\label{I6}
\varepsilon_\phi(\phi_{*})&=&\frac{2\,p^2m_\mathrm{pl}^{2p}}{v^{2p}}\biggr(\frac{v^p}{2\,p\,m_\mathrm{pl}^p((p-2)N_{*}+(p-1))}\biggr)^\frac{2(p-1)}{p-2},\\[0.4cm]
\eta_\phi(\phi_{*})&=&\frac{(1-p)}{(p-2)N_*+(p-1)}\:.
\end{eqnarray}
For our analysis, we use a typical value of $N_* = 60$. The predictions of the spectral index for models with different $p$ and for $N_*\sim 50-60$ are shown in figure \ref{fig:2}.
The value $V_0$, i.e.\ the value of the scalar potential at $\phi=0$, can be obtained with the observed value of the scalar amplitude $A_\mathrm{s}\simeq2.2\times10^{-9}$:
\begin{equation}\label{I7}
\begin{aligned}
V_0=24 \pi^2 \varepsilon_{\phi}(\phi_*)A_\mathrm{s} m_\mathrm{pl}^4\simeq  \frac{48 p^2 \pi^2 A_s m_\mathrm{pl}^{2p+4}}{v^{2p}}\biggr(\frac{v^p}{2\,p\,m_\mathrm{pl}^p((p-2)N_{*}+(p-1))}\biggr)^\frac{2(p-1)}{p-2}.\\
\end{aligned}
\end{equation}\\

\subsection{After Inflation}\label{S_AI}

\subsubsection*{Masses of the fields}
 The masses of the inflaton $m_\phi$ and the (s)neutrino $m_\chi$ after inflation, at the minimum of the potential where $\phi=v$ and $\chi=0$, are given by
\begin{eqnarray}\label{M1}
m_\phi^2\,\, &=& \frac{\partial^2 V}{\partial \phi^2}\biggr\rvert_\mathrm{min} =\frac{2 p^2\,V_0}{v^2}\:,\label{M1-1}\\
m_\chi^2 &=& \frac{\partial^2 V}{\partial \chi^2}\biggr\rvert_\mathrm{min} =\frac{4 \lambda^2}{2^n}\, v^{2n}\:.\label{M1-2}
\end{eqnarray}

\subsubsection*{Preheating: parametric resonance}\label{s_PPR}

\begin{figure}[t] %
\centering
\includegraphics[width=14cm,height=12cm,keepaspectratio]{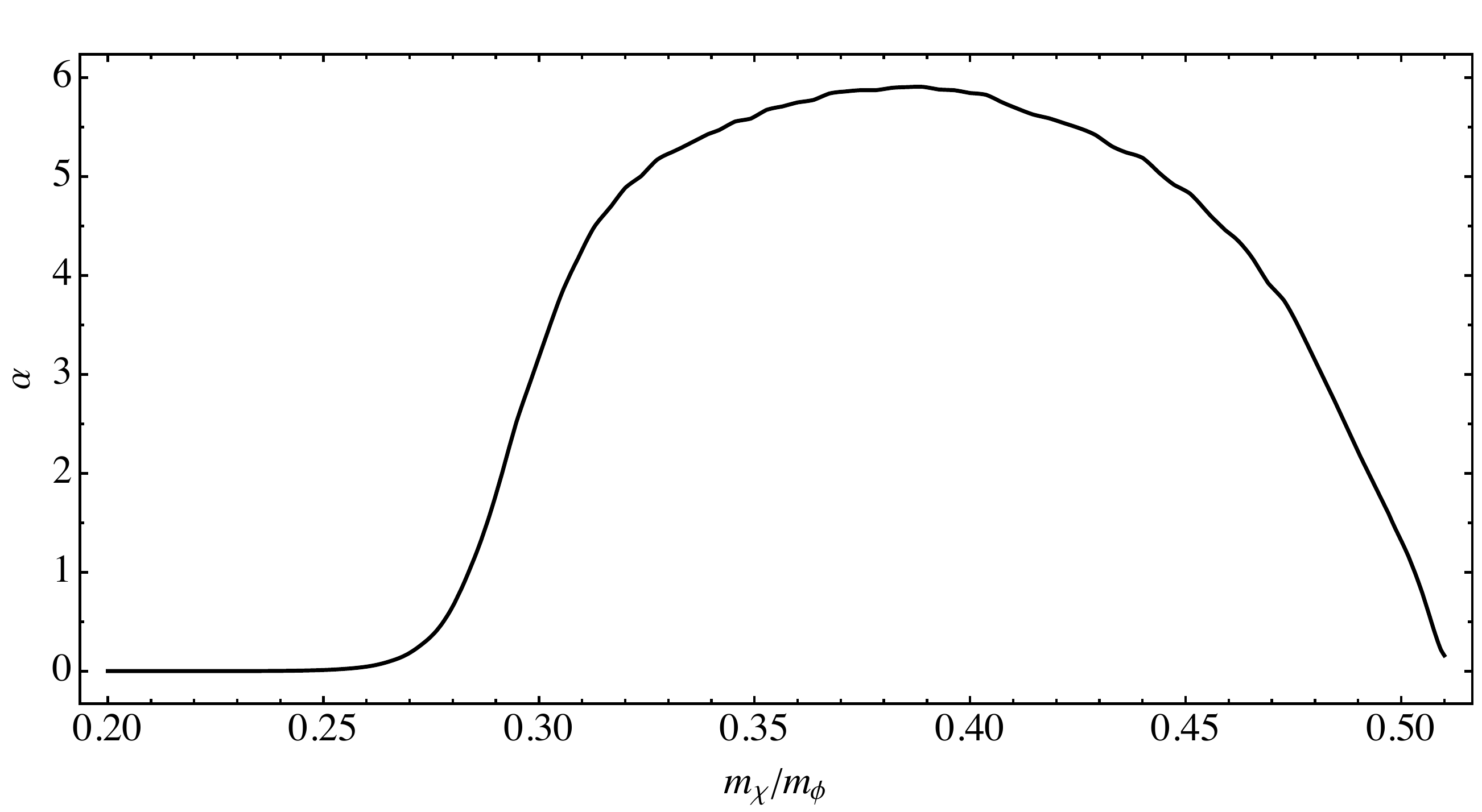}
\caption{Initial sneutrino-to-inflaton energy density ratio $\alpha=(\rho_\chi/\rho_\phi)_\mathrm{initial}$ extracted from the preheating results from \cite{Antusch:2015vna} for the $p=6$ and $v=10^{-2}m_\mathrm{pl}$ scenario.}
\label{fig:3}
\end{figure}

In the early reheating phase of the models, the fluctuations of the inflaton field grow exponentially during so-called tachyonic oscillations (cf.\ \cite{Antusch:2015nla,Brax:2010ai}). In addition, fluctuations of secondary scalar fields coupled to the inflaton can get produced via parametric resonance \cite{Kofman:1994rk,Kofman:1997yn} when the inflaton oscillates around its minimum. In \cite{Antusch:2015vna} this resonance has been studied using lattice simulations (in 2D) for the example model with $p=6$ and $v=10^{-2}\,m_\mathrm{pl}$. A strong resonance band has been found for $0.25<m_\chi/m_\phi \lesssim 0.5$ (cf.\ figure \ref{fig:3}). A similar resonance can be found for $p=4$ and other values of $v$. 

In the following, we will use the result from \cite{Antusch:2015vna} (with $p=6$ and $v=10^{-2}\,m_\mathrm{pl}$) as a first estimate for 
the initial sneutrino-to-inflaton energy density ratio $\alpha$ at the beginning of perturbative reheating, defined as 
\begin{equation}\label{PR1}
\alpha = \frac{\rho_\chi}{\rho_\phi}\biggr\rvert_\mathrm{initial},\\
\end{equation}
for all considered $p$ and $v$. The result for $\alpha$ from the lattice simulation of \cite{Antusch:2015vna} is shown in figure \ref{fig:3}, where we have used that the variances of the fields are related to the mean energy densities via $\langle\rho_f\rangle=m_f^2\langle \delta f^2\rangle$.
For the future, it would be desirable to have results from 3D lattice simulations available for all $p$ values and $v$ ranges, however this is very computationally expensive and beyond the scope of this paper. 

Using $\alpha $ one can express the initial inflaton and (s)neutrino energy densities in terms of the total energy density as
\begin{eqnarray}\label{PR2}
\rho_\phi\,\, &=& \frac{1}{1+\alpha}\,\rho_\mathrm{tot},\\[0.5cm]
\rho_\chi \,\, &=& \frac{\alpha}{1+\alpha}\,\rho_\mathrm{tot},
\end{eqnarray}
where $\rho_\mathrm{tot}$ is related to the Hubble parameter by $3m_\mathrm{pl}^2 H^2=\sum_{i}\rho_i=\rho_\mathrm{tot}$.

\subsubsection*{Perturbative decay of the inflaton and the (s)neutrino}\label{}
During the phase of perturbative reheating the inflaton decays into right-handed (s)neutrinos ($\phi  \rightarrow \chi\, \chi$, $\phi  \rightarrow \overline \chi\,  \overline \chi$ and $\phi  \rightarrow \psi_\chi\, \psi_\chi$) and the (s)neutrinos then decay into MSSM particles.\footnote{As a complex scalar field, the sneutrino has two real components, which we denote here by $\chi$ and $\overline{\chi}$.} Assuming $m_\phi>2m_\chi$ for the lightest right-handed (s)neutrino mass and $m_\chi \gg m_{L_j}, m_{H_u}$, this process is kinematically allowed and the decay rates of the inflaton and the lightest right-handed (s)neutrinos are named $\Gamma_\phi$ and $\Gamma_\chi$, respectively. To calculate the inflaton decay rate $\Gamma_\phi$ we expand the scalar potential of Eq.~(\ref{B2}) around the vacuum expectation value $v$ of $\phi$ in the minimum after inflation, i.e.\ $\phi=v+\delta \phi$ with small perturbations $\delta \phi$, 
 and include also the fermionic interactions from the superpotential of Eq.~(\ref{B1}). We obtain:
\begin{equation}\label{R1}
\Gamma_{\phi}=\frac{1}{4\pi}\biggr(\biggr(\frac{2 \lambda}{m_\chi}\biggr)^{1/n}\frac{n \, m_\chi}{4}\biggr)^2 m_\phi \biggr(\,1+12\frac{m^2_\chi}{m^2_\phi}\,\biggr)\sqrt{1-4\frac{m^2_\chi}{m^2_\phi}}.
\end{equation}
Details on the derivation of $\Gamma_\phi$ are given in appendix \ref{InflatonDecay}. 
The right-handed (s)neutrinos decay further into Higgs(ino)s and (s)leptons (which are treated as massless in our analysis) via their Yukawa couplings, and we assume here that the decay products thermalise quickly.
The (s)neutrino decay rate $\Gamma_\chi$ is given by (cf.\ e.g.\ \cite{Antusch:2014qqa}):
\begin{equation}\label{R2}
\Gamma_{\chi}=\Gamma_{\overline \chi}=\Gamma_{\psi_{\chi}}=\frac{m_\chi}{4 \pi} \sum_{j} |y_{j}|^2.
\end{equation}
In the following, defining $y^2 \equiv \sum_{j} |y_{j}|^2$, we will use the estimate
\begin{equation}\label{R2}
m_{\nu} \sim \, \frac{y^2 \langle h_u\rangle^2}{m_\chi}
\end{equation}
for the light neutrino mass generated by the lightest right-handed neutrino via the seesaw mechanism. 
Plugging in $\langle h_u\rangle = 174$ GeV gives:
\begin{equation}\label{R2}
y^2 \sim \, \biggr(\frac{m_{\nu}}{1\, \text{meV}}\biggr)\biggr(\frac{m_{\chi}}{3\times 10^{16}\,\text{GeV}}\biggr)  \:.
\end{equation}
The reheat temperature $T_\mathrm{RH}$ represents the temperature when the first radiation dominated phase after inflation begins, i.e.\ when most of the inflatons have decayed into MSSM particles. 
For the case that $\Gamma_\chi > \Gamma_\phi$ it can be estimated by the following expression (see e.g.\ \cite{Antusch:2010mv}): 
\begin{equation}\label{R4}
\begin{aligned}
T_\mathrm{RH}&=\biggr(\frac{36}{g_* \pi^2}m_\mathrm{pl}^2 \Gamma_\phi^2\biggr)^{\frac{1}{4}}\\&=\biggr(\frac{3n^2 m_\mathrm{pl}}{32 \pi^2\sqrt{g_*}} \biggr(\frac{2 \lambda}{m_\chi}\biggr)^{{\frac{2}{n}}} m_\chi^2 m_\phi \biggr)^{\frac{1}{2}} \biggr(1+12 \frac{m_\chi^2}{m_\phi^2}\biggr)^{\frac{1}{2}}\biggr(1-4 \frac{m_\chi^2}{m_\phi^2}\biggr)^{\frac{1}{4}},
\end{aligned}
\end{equation}
which is obtained via an approximate analytical solution of a set of Boltzmann equations and the assumption of matter domination, i.e.\ $\Hubble \simeq \tfrac{2}{3}t^{-1}$. The total number of relativistic degrees of freedom is given by $g_*=915/4$ in case of the MSSM \cite{Davidson:2008bu}. To ensure that leptogenesis is indeed non-thermal, the condition $m_\chi \gg T_\mathrm{RH}$ will be required. In the framework of supergravity, non-thermal leptogenesis has the advantage that it can work with lower $T_\mathrm{RH}$, which is better compatible with bounds from avoiding overproduction of gravitinos \cite{Khlopov:1984pf,Ellis:1984eq,Moroi:1993mb,Kawasaki:2004yh,Hook:2018sai}, which typically imposes $T_\mathrm{RH} \lesssim 10^{7}-10^{10}$ GeV (depending on the specific model).\\

\section{Calculation of the Baryon Asymmetry}\label{S2}

\subsection*{Boltzmann equations}

The perturbative part of the reheating process can be described by the following set of Boltzmann equations:
\begin{eqnarray}\label{BE1}
\dot{n}_\phi(t)+3H(t)n_\phi(t)+\Gamma_{\phi}n_\phi(t)&=&0,\label{BE1-1}\\[0.2cm]
\dot{n}_\chi(t)+3H(t)n_\chi(t)-2\,\Gamma_\phi n_\phi(t)+\Gamma_\chi n_\chi(t)&=&0,\label{BE1-2}\\
\dot{\rho}_{rad}(t)+4H(t)\rho_{rad}(t)-\Gamma_\chi m_\chi n_\chi(t)-\biggr(\,1-2\,\frac{m_\chi}{m_\phi}\,\biggr)\,\Gamma_\phi\,m_\phi n_\phi(t)&=&0,\label{BE1-3}
\end{eqnarray}
where for a particle at rest, such as the inflaton, $\rho_\phi=n_\phi m_\phi$ accounts. This set of equations describes the evolution of the number densities of the inflaton ($n_\phi$) and the sneutrino ($n_\chi$), and the radiation energy density $\rho_{rad}$, which covers the final decay products, i.e.\ the (s)leptons and Higgs(ino)s and the particles produced from their decays. 

The last term in eq.\ (\ref{BE1-3}) takes into account that the inflaton can decay into a (s)neutrino with a significant amount of kinetic energy, depending on the mass ratio $m_\chi/m_\phi$. The kinetic energy then red-shifts like radiation and is counted as part of $\rho_{rad}$. The Hubble parameter is determined by the first Friedmann equation:
\begin{equation}\label{BE2}
H^2= \frac{1}{3 m_\mathrm{pl}^2}\biggr(n_\phi m_\phi+n_\chi m_\chi+\rho_{rad}\biggr).
\end{equation}
The parameter $\alpha$ which takes the effects of parametric resonance into account, introduced in eq.\ (\ref{PR1}), sets the initial conditions for $\rho_\phi=n_\phi m_\phi$ and $\rho_\chi=n_\chi m_\chi$. The inflaton decay gets efficient when $H \sim \Gamma_\phi$. The sneutrino abundance present due to parametric resonance may already start to decay earlier at $H \sim \Gamma_\chi$ if $\Gamma_\chi> \Gamma_\phi$. The evolution of the effective lepton number density $n_L$ is covered by an additional Boltzmann equation \cite{Antusch:2010mv}:
\begin{equation}\label{BE3}
\dot{n}_L(t)+3H(t)\,n_L(t)=\epsilon\,\Gamma_\chi n_\chi(t),
\end{equation}
where $\epsilon$ represents the CP-violation per sneutrino decay. In the following, we will use the upper bound for the right-handed (s)neutrino decay asymmetry from \cite{Covi:1996wh,Davidson:2002qv,Hamaguchi:2001gw},
\begin{equation}\label{BE4}
\begin{aligned}
\epsilon<\frac{3}{8\,\pi}\frac{\sqrt{\Delta m_\mathrm{atm}^2}\,m_\chi}{\langle h_u\rangle^2},
\end{aligned}
\end{equation}
for calculating the maximally produced baryon asymmetry. The measured value of the mass squared difference is $\Delta m_\mathrm{atm}^2 \simeq 2.5\times 10^{-3}\,\text{eV}^2$ and $\langle h_u\rangle=174\,\text{GeV}$. 


\subsection*{Baryon asymmetry}

The maximally produced lepton asymmetry is converted into a baryon asymmetry by sphaleron processes such that $n_B=C/(C-1)\,n_L$ with $C=1/3$ for the MSSM (cf.\ \cite{Davidson:2008bu}). A normalisation by the photon number density is common, which is deduced from the entropy density $s=7.04\,n_\gamma$:
\begin{equation}\label{BA1}
\begin{aligned}
\biggr\rvert\frac{n_B}{n_\gamma}\biggr\lvert=7.04\, \biggr\rvert\frac{C}{C-1}\frac{n_L}{s}\biggr\lvert.\\
\end{aligned}
\end{equation}
The entropy density $s=2\pi^2 g_* T^3/45$ is given by the temperature $T^4=30\,\rho_{rad}/g_* \pi^2$, which in return is obtained from the radiation energy density in eq.\ (\ref{BE1-3}).

\subsection*{Approximation via $T_\mathrm{RH}$ and dependency of $n_B/n_\gamma$ on $n$, $v$ and $m_\chi$}

An analytic estimate for the baryon asymmetry can be given in terms of the reheat temperature (see e.g.\ \cite{Antusch:2010mv}):
\begin{equation}\label{BA2}
\begin{aligned}
\biggr\rvert\frac{n_B}{n_\gamma}\biggr\lvert_\mathrm{approx}=8.8\frac{\epsilon \, T_\mathrm{RH}}{m_\phi},
\end{aligned}
\end{equation}
which is obtained via the analytic solution of a simplified Boltzmann equation used already for the reheat temperature where matter domination has been assumed. 

For $\Gamma_\chi \gg \Gamma_\phi$, using the reheat temperature estimate from eq.\ (\ref{R4}) and plugging in the form of the decay rate $\Gamma_\phi$ given in eq.\ (\ref{A6}) as well as the inflaton mass from eq.~(\ref{M1-1}), the dependencies of the baryon asymmetry on the parameters $n$, $p$, $v$ and $m_\chi$ can be extracted. For example, one can see that (in this approximation) the produced baryon asymmetry depends linearly on $n$:
\begin{equation}\label{BA3}
\begin{aligned}
\biggr\rvert\frac{n_B}{n_\gamma}\biggr\lvert_\mathrm{approx}=8.8\,\epsilon\frac{n\,m_\chi}{v\,\sqrt{m_\phi}}\biggr(\frac{3 m_\mathrm{pl}}{16\pi^2\sqrt{g_*}}\biggr(1+12\frac{m_\chi^2}{m_\phi^2}\biggr)\sqrt{1-4\frac{m_\chi^2}{m_\phi^2}}\biggr)^{1/2}.
\end{aligned}
\end{equation}
These approximate dependencies are also confirmed by our results using Boltzmann equations.

\section{Results for the Baryon Asymmetry}\label{S3}

\subsection*{Importance of including the effects of preheating}
\begin{figure}[t]
\centering
\includegraphics[width=13cm,height=10cm,keepaspectratio]{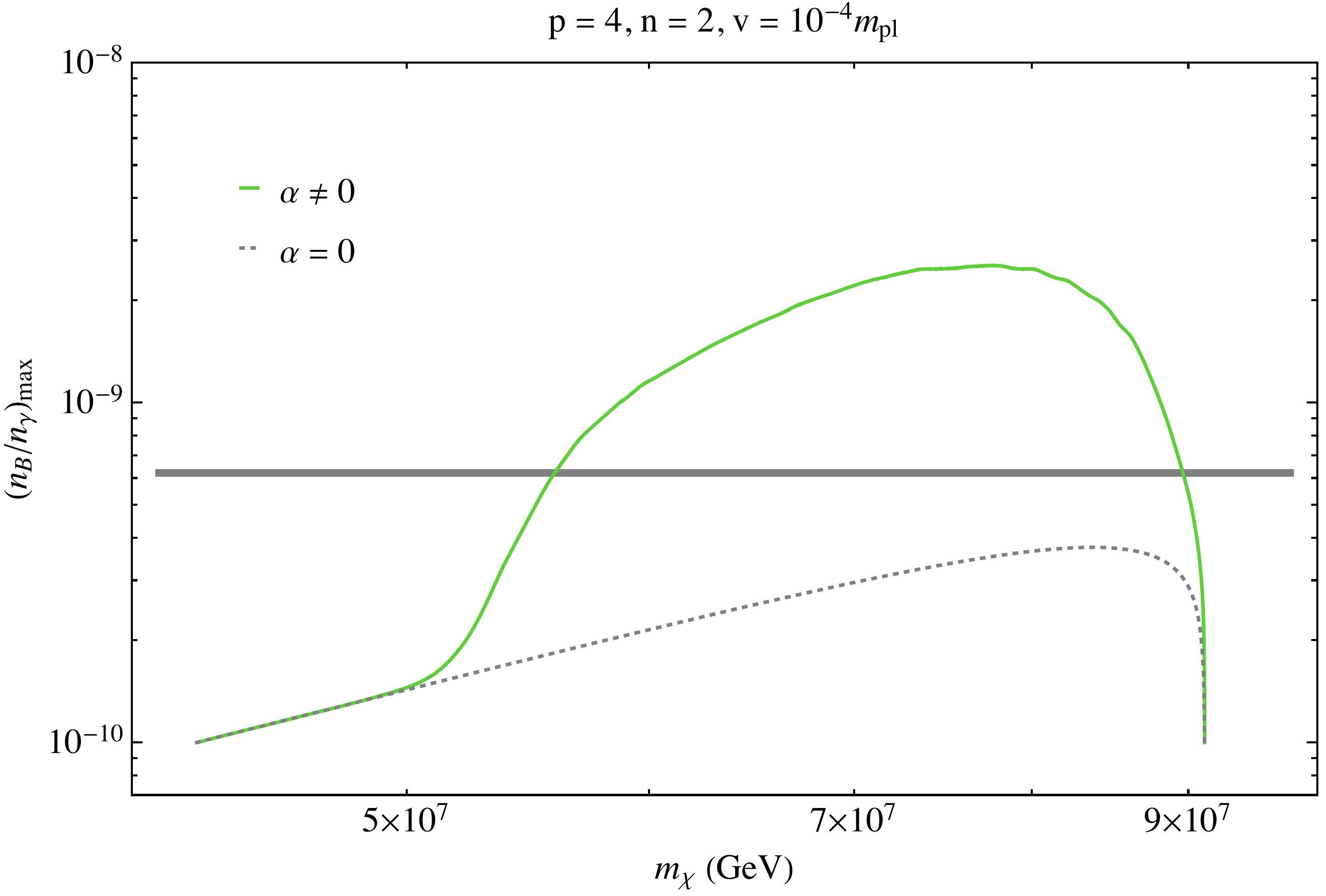} 
\caption{Maximally produced baryon asymmetry for $n=2$, $p=4$ and $v=10^{-4}m_\mathrm{pl}$. The grey dotted line labeled with $\alpha=0$ indicates $(n_B/n_\gamma)_\mathrm{max}$ where preheating effects have been ignored and $\alpha \not=0$ marks the results which include the preheating effects. The horizontal grey bar indicates the value of observed baryon asymmetry $(n_B/n_\gamma)=(6.19\pm 0.15)\times 10^{-10}$.}
\label{fig:v4n2p4}
\end{figure}

To illustrate the importance of including the effects of preheating, we show an example result of $(n_B/n_\gamma)_\mathrm{max}$ in figure \ref{fig:v4n2p4},  obtained from solving the Boltzmann equations discussed in section (\ref{S_AI}). Two cases are shown, one where preheating is ignored, labeled as $\alpha = 0$, and one with $\alpha \not= 0$ where preheating effects are included as described in section \ref{s_PPR}. In the example we have taken the model parameters to be $p=4$, $n=2$, $v= 10^{-4}\:m_\mathrm{pl}$ and $m_\nu=1\,\text{meV}$.

We find that preheating effects can enhance the produced asymmetry by almost an order of magnitude. In the present example, without them it would not be possible to explain the observed value of $(n_B/n_\gamma)=(6.19\pm 0.15)\times 10^{-10}$ which is indicated by the grey bar. Only when they are included, a sufficient baryon asymmetry can be generated within a window for the right-handed (s)neutrino mass of $5.59 \times 10^{7}$ to $8.96 \times 10^{7}$ GeV.

\begin{figure}%
\centering
\subfigure{%
\includegraphics[height=6.7cm]{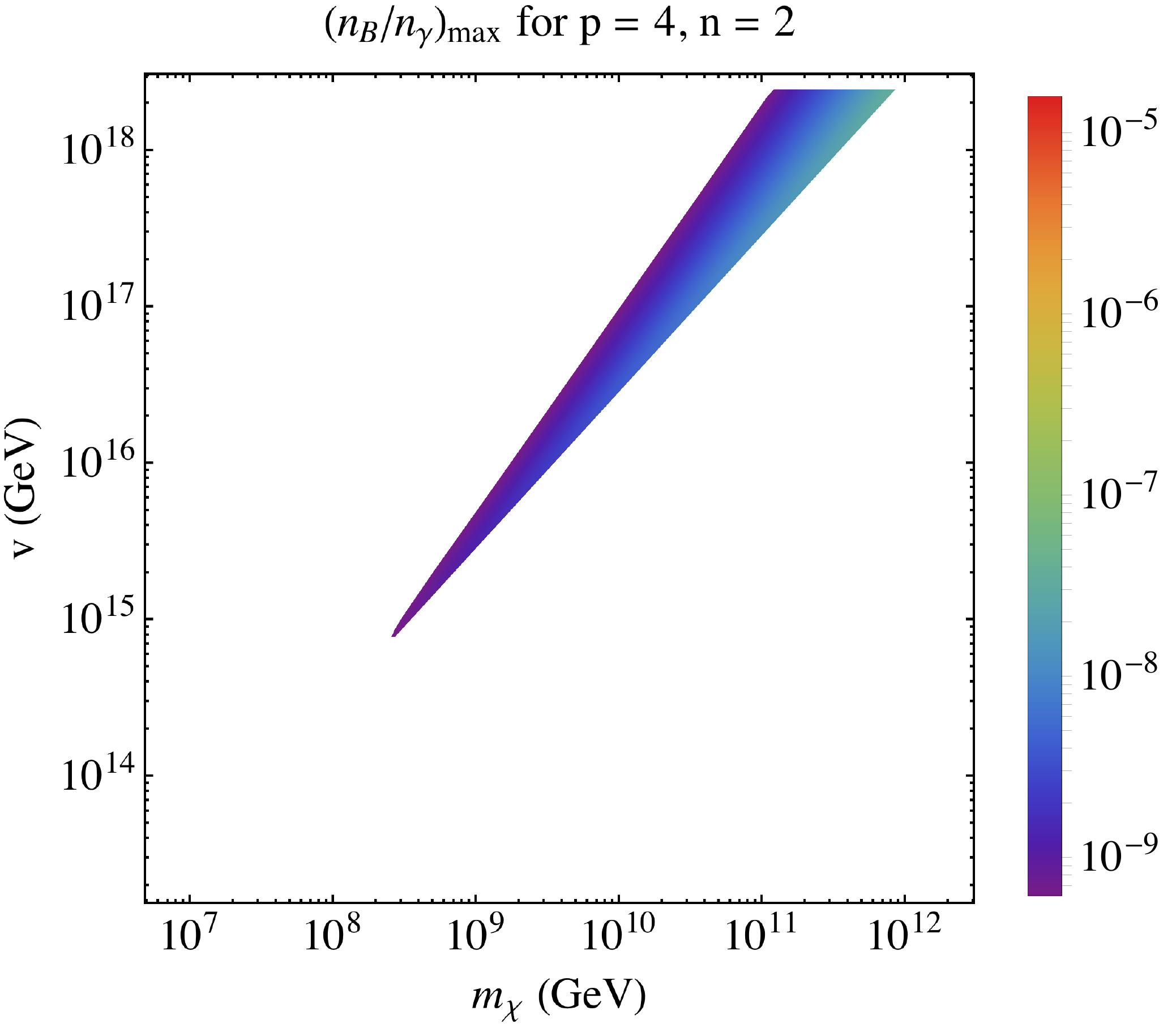}}%
\hspace{0pt}
\subfigure{%
\includegraphics[height=6.7cm]{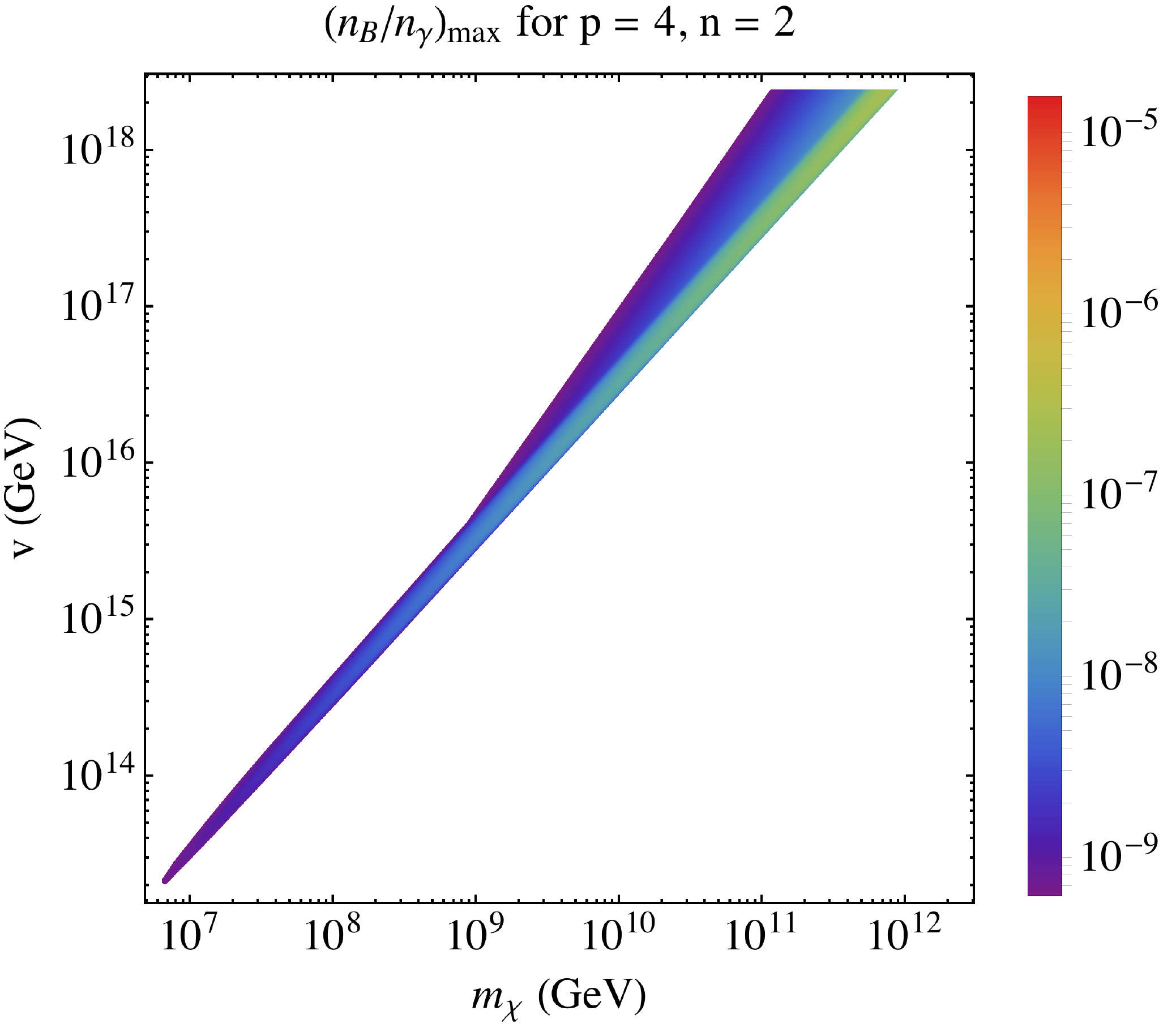}}
\caption{Maximally produced baryon asymmetry as a function of the right-handed (s)neutrino mass $m_\chi$ and vacuum expectation value $v$ of the inflaton field after inflation, for model parameters $n=2$ and $p=4$. On the left preheating effects were ignored, whereas on the right they have been included. }
\label{fig:nB_p4_n2}
\end{figure}

The maximally produced baryon asymmetry is shown again in figure \ref{fig:nB_p4_n2} for parameters $p=4$ and $n=2$ as a function of $v$ and $m_\chi$. On the left the effects of preheating have been ignored for the evaluation of $(n_B/n_\gamma)_\mathrm{max}$, whereas on the right they have been included. The effect of preheating can be seen by the narrow band of enhanced baryon asymmetry stretching beyond the triangle visible in the left figure, i.e.\ preheating may lead to consistent $(n_B/n_\gamma)_\mathrm{max}$ for much smaller values of $v$ and $m_\chi$.

\subsubsection*{Comparison of produced baryon asymmetry for different model parameters}

\begin{figure}[t]%
\centering
\subfigure{%
\includegraphics[height=7.8cm]{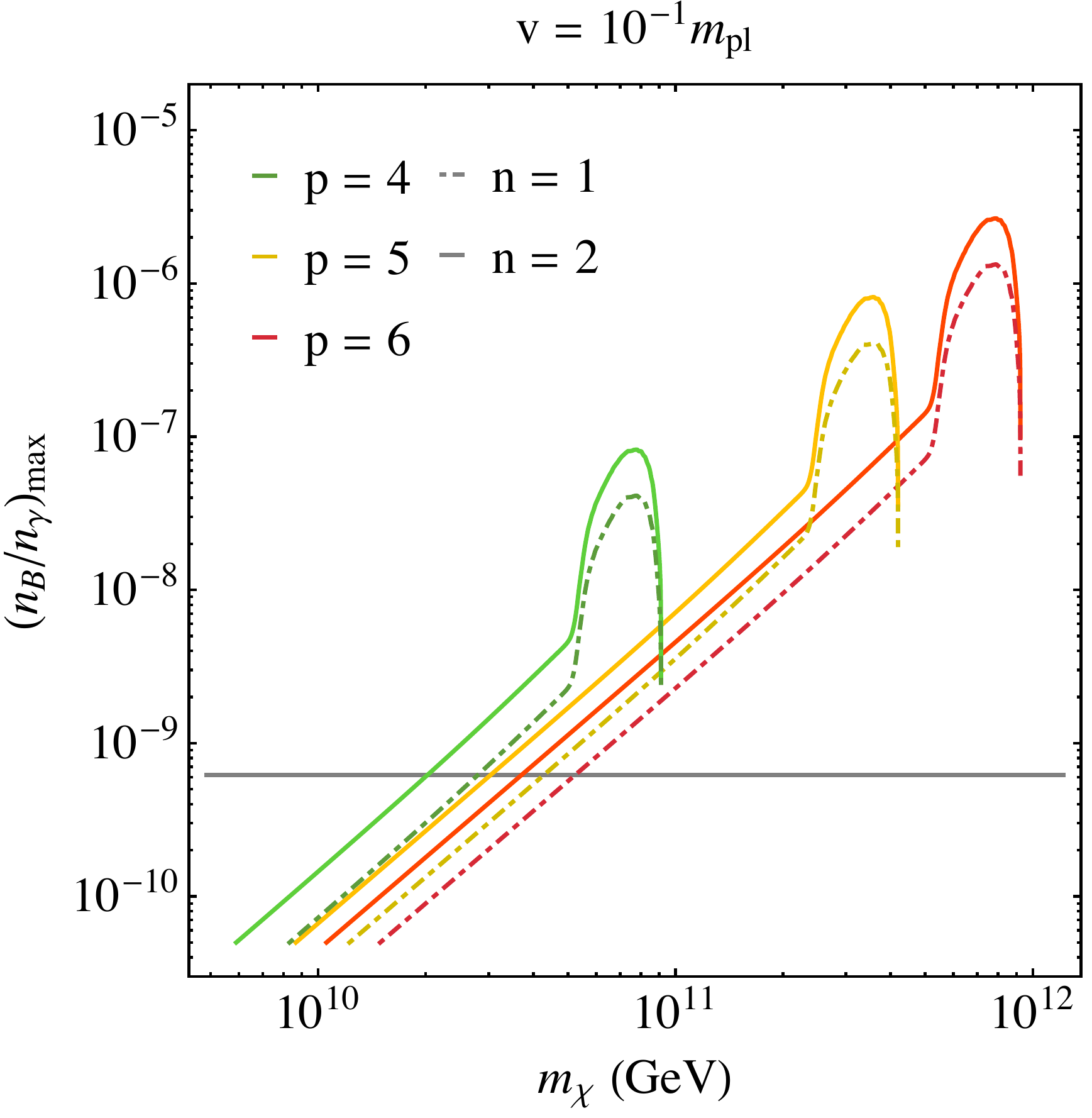}}%
\hspace{0pt}
\subfigure{%
\includegraphics[height=7.8cm]{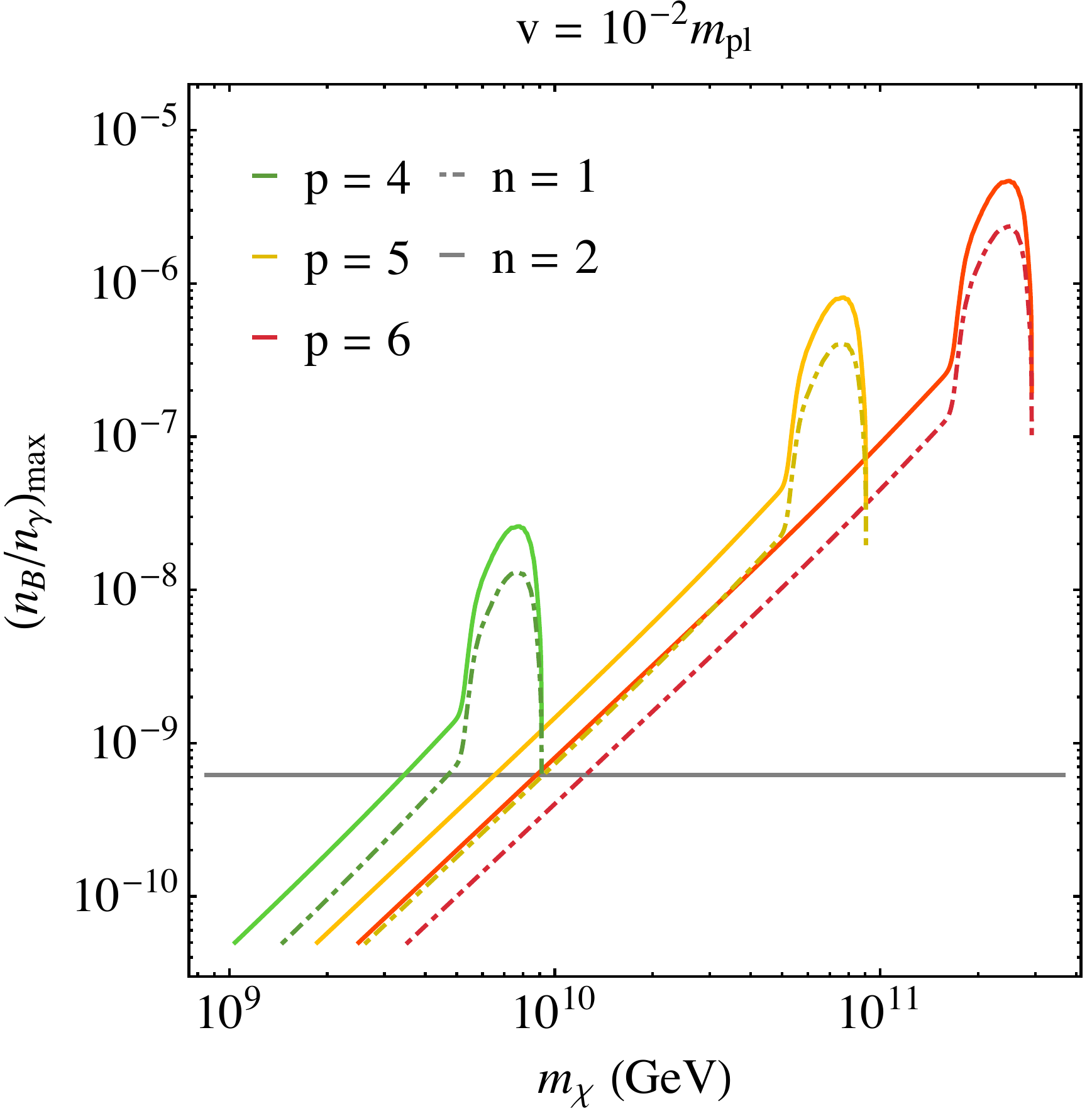}}%
\hspace{0pt}
\subfigure{%
\includegraphics[height=7.8cm]{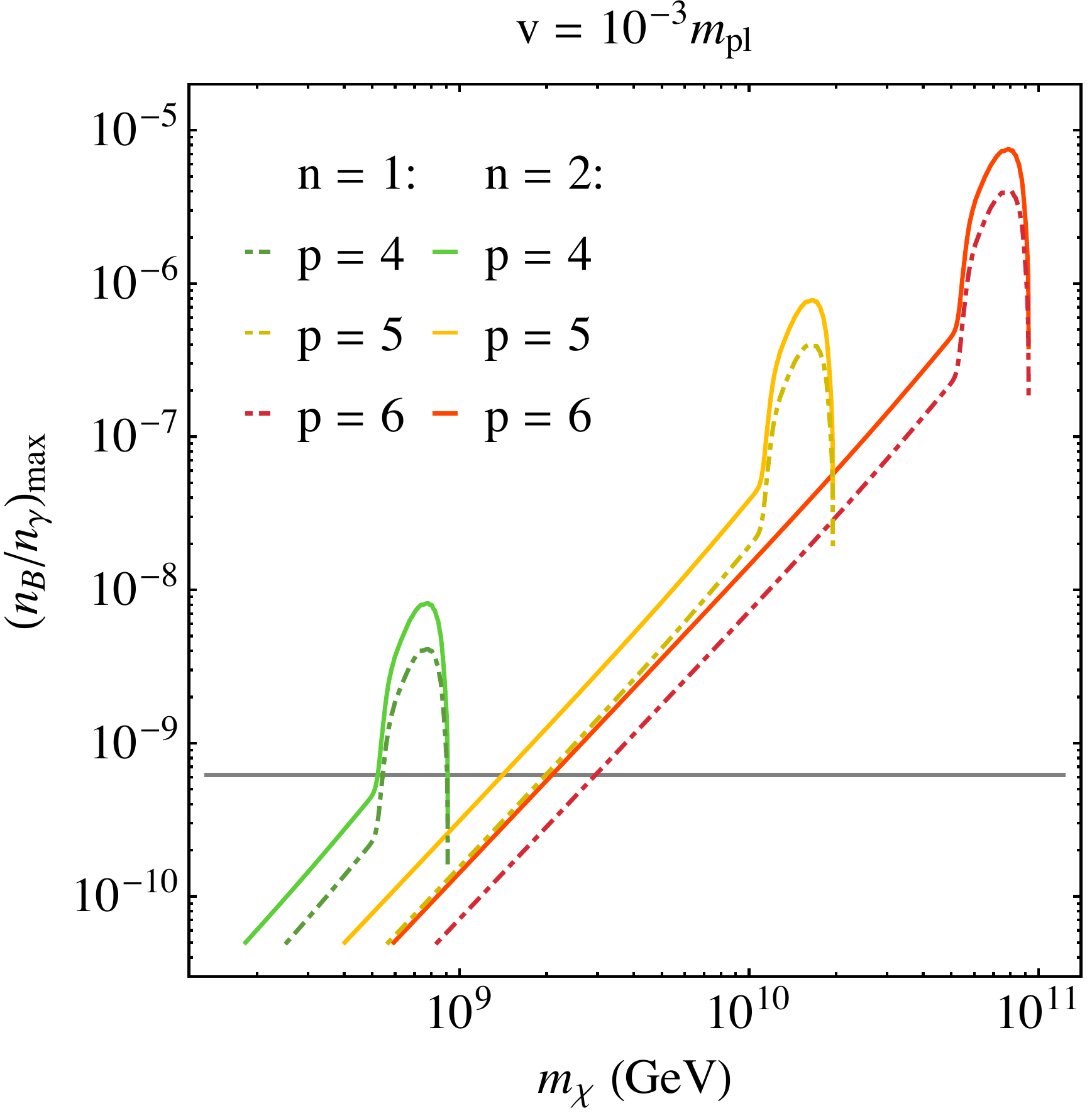}}%
\hspace{0pt}
\subfigure{%
\includegraphics[height=7.8cm]{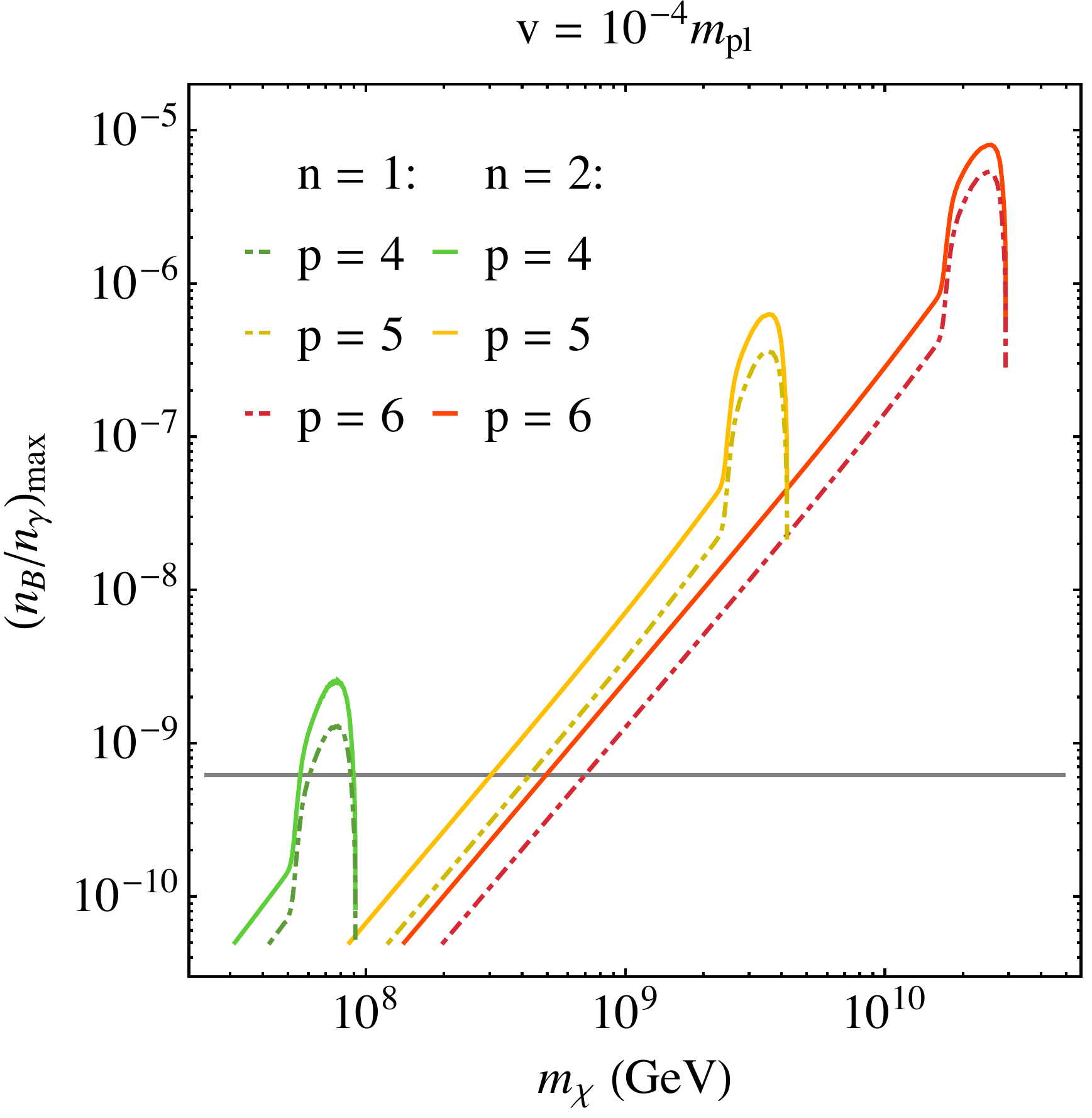}}%
\caption{Maximally produced baryon asymmetry as a function of $m_\chi$ with fixed model parameters $n=1, 2$ and $p=4$ (green), $5$ (yellow), $6$ (red) for different vacuum expectation values $v=10^{-1},10^{-2},10^{-3}$ and $10^{-4}m_\mathrm{pl}$. The case $n=1$ is indicated by the coloured dash-dotted and $n=2$ by the solid lines. The grey horizontal line indicates the observed value of $n_B/n_\gamma=6.19\times10^{-10}$ such that larger values of $(n_B/n_\gamma)_\mathrm{max}$ can be consistent with observations (with $\epsilon <  \epsilon_\mathrm{max}$). The effect of preheating appears as an enhanced asymmetry visible by the bulge on the right end.}
\label{fig:BE_Pre_p456}
\end{figure}

To demonstrate the differences between the maximally produced baryon asymmetry in the models with $n=1,2$ and $p=4,5,6$, we show $(n_B/n_\gamma)_\mathrm{max}$ as a function of $m_\chi$ in figure \ref{fig:BE_Pre_p456} for different vacuum expectation values $v$. The light (``left-handed'')  neutrino mass, used instead of the Yukawa coupling (see eq.~(\ref{R2})), is set to $m_\nu=1\,\text{meV}$. We note that unless $m_\nu$ is chosen much smaller than $1\,\text{meV}$, our results are insensitive to $m_\nu$. 

The grey horizontal line indicates the observed value of $n_B/n_\gamma=(6.19\pm 0.15)\times 10^{-10}$. For model parameters which give a value of $n_B/n_\gamma$ below this line it is not possible to produce sufficient baryon asymmetry via the non-thermal leptogenesis mechanism. Since we have used $\epsilon_\mathrm{max}$ for calculating the produced baryon asymmetry, model parameters which give values of $n_B/n_\gamma$ above the grey line can also be consistent (when the actual decay asymmetry satisfies $\epsilon <  \epsilon_\mathrm{max}$).

The effect of preheating can be seen by the sudden enhancement of $(n_B/n_\gamma)_\mathrm{max}$ on the right side of the plots. Without preheating the asymmetry would continue to grow approximately linearly on the logarithmic scale (as the grey dotted line in figure \ref{fig:v4n2p4}). A general feature of the results is that smaller values of $m_\chi$ for fixed $n$ and $p$ lead to a smaller maximal asymmetry. Comparing $n=2$ with $n=1$ the asymmetry is increased. From the approximate formula in eq.~(\ref{BA3}) one can see that $(n_B/n_\gamma)_\mathrm{max}$ is larger by a factor of $\approx 2$ for $n=2$ compared to $n=1$. 

The result for the different parameters $p$ show that for fixed values of $v$ and $m_\chi$ the asymmetry becomes larger for smaller $p$. On the other hand, increasing $p$ allows for larger possible values of $m_\chi$, which in turn increases $(n_B/n_\gamma)_\mathrm{max}$. The allowed ranges for the values of $m_\chi$ can be found in table \ref{tab:1} for the example values $v=10^{-2}m_\mathrm{pl}$ (where $v$ is of the order of the Grand Unification scale) and $v=10^{-4}m_\mathrm{pl}$. The $v=10^{-4}m_\mathrm{pl}$ case includes the results depicted in figure \ref{fig:v4n2p4} and shows that also for $p=4$, $n=1$, due to the preheating effects, sufficient baryon asymmetry can be produced.

\begin{table}
  \label{tab:1}
  \begin{center}
    \begin{tabular}{| *6{>{\centering\arraybackslash}m{1.9cm}|} @{}m{0pt}@{}}
    \hline
    \multicolumn{2}{ |c| }{} & \multicolumn{2}{ c }{\textbf{$m_\chi(v=10^{-2}m_\mathrm{pl})$}} & \multicolumn{2}{ |c| }{\textbf{$m_\chi(v=10^{-4}m_\mathrm{pl})$}} &\\[2ex] 
    \hline
    & \textbf{$p = 4$} & \multicolumn{2}{ c }{$4.69\times10^9-9.11\times10^9\, \text{GeV}$} & \multicolumn{2}{ |c| }{$6.07\times10^7-8.73\times10^7\, \text{GeV}$} &\\[2ex]
    \cline{2-6}
    \textbf{$n = 1:$} & \textbf{$p = 5$} & \multicolumn{2}{ c }{$9.22\times10^9-9.05\times10^{10}\, \text{GeV}$} & \multicolumn{2}{ |c| }{$4.28\times10^8-4.2\times10^{9}\, \text{GeV}$} &\\[2ex]
    \cline{2-6}
     & \textbf{$p = 6$} & \multicolumn{2}{ c }{$1.24\times10^{10}-2.92\times10^{11}\, \text{GeV}$} & \multicolumn{2}{ |c| }{$6.99\times10^{8}-2.92\times10^{10}\, \text{GeV}$} &\\[2ex]
    \hline
     & \textbf{$p = 4$} & \multicolumn{2}{ c }{$3.45\times10^9-9.11\times10^9\, \text{GeV}$} & \multicolumn{2}{ |c| }{$5.59\times10^7-8.96\times10^7\, \text{GeV}$} &\\[2ex]
    \cline{2-6}
    \textbf{$n = 2:$} & \textbf{$p = 5$} & \multicolumn{2}{ c }{$6.54\times10^9-9.05\times10^{10}\, \text{GeV}$} & \multicolumn{2}{ |c| }{$3.04\times10^8-4.2\times10^{9}\, \text{GeV}$} &\\[2ex]
    \cline{2-6}
     & \textbf{$p = 6$} & \multicolumn{2}{ c }{$8.79\times10^9-2.92\times10^{11}\, \text{GeV}$} & \multicolumn{2}{ |c| }{$4.95\times10^8-2.92\times10^{10}\, \text{GeV}$} &\\[2ex]
    \hline
    \end{tabular}
  \end{center}
  \caption{Allowed ranges of $m_\chi$ compatible with the observed $n_B/n_\gamma$ for model parameters $n=1,2$, $p=4,5,6$ and the example values $v=10^{-2}m_\mathrm{pl}$ (where $v$ is of the order of the Grand Unification scale) and $10^{-4}m_\mathrm{pl}$.}
\end{table}

\subsection*{Baryon asymmetry and $T_\mathrm{RH}$ in the $v$-$m_\chi$ plane}
As we have discussed above, the cases with $n=1$ and $n=2$ only differ by a global factor of approximately $2$. We therefore focus in this section on the case $n=2$. We show our results for $(n_B/n_\gamma)_{\mathrm{max}}$ and the estimated $T_\mathrm{RH}$ as a function of $m_\chi$ and $v$ for $p=4,5,6$ in figure \ref{fig:nB_Trh}. The results on the left generalise our plots from figure \ref{fig:BE_Pre_p456} for general $v$. As in the previous section the left-handed neutrino mass is used instead of the Yukawa coupling and is set to $m_\nu=1\,\text{meV}$ where larger values do not affect the results. In addition, we have included a white/black dashed line for the $p=5,6$ cases which indicates $m_\chi=10\times T_\mathrm{RH}$. Above the line the thermal production and washout processes can be neglected. In the region far below the dashed line thermal effects cannot be ignored and our treatment can not be applied. For the shown values of $v$,  $m_\chi > T_\mathrm{RH}$ still holds, but the results below the dashed lines should be taken with some caution. On the right of the coloured region where non-thermal leptogenesis works the decays of $\phi$ into two right-handed (s)neutrinos is kinematically forbidden ($m_\phi < 2m_\chi$) and on the left it is not possible to generate the observed value $(n_B/n_\gamma)=6.1\times10^{-10}$ even with a maximal decay asymmetry $\epsilon_\mathrm{max}$.

The effects of preheating on the $(n_B/n_\gamma)_\mathrm{max}$ results is mainly visible by the small band of enhanced asymmetry close to the $m_\phi=2m_\chi$ border. The importance of them has been discussed for $p=4$, $n=2$ within figure \ref{fig:nB_p4_n2}. For $p=5$ and $6$, preheating enhances $(n_B/n_\gamma)_\mathrm{max}$ as well, but unlike for $p=4$ only for values of $v$ and $m_\chi$ where the generation of sufficient baryon asymmetry would have been possible also without preheating effects. 
On the right we show the estimates for $T_\mathrm{RH}$ using the approximation from section \ref{S_AI}. The results for $T_\mathrm{RH}$ spread over several orders of magnitude for the different cases. The $p=4$ case shows that preheating not only enhances $(n_B/n_\gamma)_{\mathrm{max}}$ but may lead to much smaller $T_\mathrm{RH}$ as well. In fact the lowest possible reheat temperatures for all model parameters can be achieved here and lies at $T_\mathrm{RH}\simeq 2\times10^5\,\text{GeV}$.

\section{Summary and Conclusions}

In this paper we have investigated how non-thermal leptogenesis can be realised in models of Majoron hilltop inflation via the decays of right-handed (s)neutrinos into (s)leptons and Higgs(ino)s during the reheating process. For that purpose we have numerically solved the Boltzmann equations for the number densities of the inflaton and the (s)neutrinos, the effective lepton number density and the radiation-like energy density. For the CP-violation in the (s)neutrino decay we have used an upper bound leading to the maximally possible lepton asymmetry. Effects of preheating, namely parametric resonance, have been included and studied for the process of non-thermal leptogenesis. The lepton asymmetry is converted into the baryon asymmetry via sphaleron processes and we presented the results for the maximally possible baryon asymmetry as functions of the vacuum expectation value and right-handed (s)neutrino mass for the different model parameters. Furthermore, we have estimated the reheat temperature $T_\mathrm{RH}$ for the different example models and studied how the required $T_\mathrm{RH}$ is influenced by the preheating effects.

We have found that the results for the maximal baryon asymmetry are strongly influenced in a variety of ways by the different model parameters and the preheating effects: The parameter $n$ from the model (\ref{B1}) appearing in the inflaton to (s)neutrino coupling turns out to have an approximately linear effect on the produced baryon asymmetry when all other model parameters are kept fixed, i.e.\ for $n=2$ the baryon asymmetry is twice as large as for $n=1$. The parameter $p$ has a more complicated dependency, but if all other model parameters are kept fixed the maximal baryon asymmetry is larger for smaller $p$, although the largest possible values appear for larger $p$. Furthermore, the allowed ranges of the right-handed (s)neutrino mass get larger for larger $p$ (c.f.\ table \ref{tab:1}). Preheating effects can enhance the produced baryon asymmetry by about an order of magnitude. Due to preheating effects, for the case $p=4$, also smaller  values of $v$ and $m_\chi$ can produce sufficient baryon asymmetry (which would have been excluded without preheating effects). The preheating effects also have an impact on the required reheat temperature, which can now be significantly lower than previously estimated.

\section*{Acknowledgements}
This work has been supported by the Swiss National Science Foundation. We thank Francesco Cefala for valuable comments on the manuscript.

\appendix

\vspace{0.5cm}
\section*{Appendix}
\section{Inflaton Decay Rate}\label{InflatonDecay}

\begin{figure}[t] 
\centering
\includegraphics[width=15cm,height=14cm,keepaspectratio]{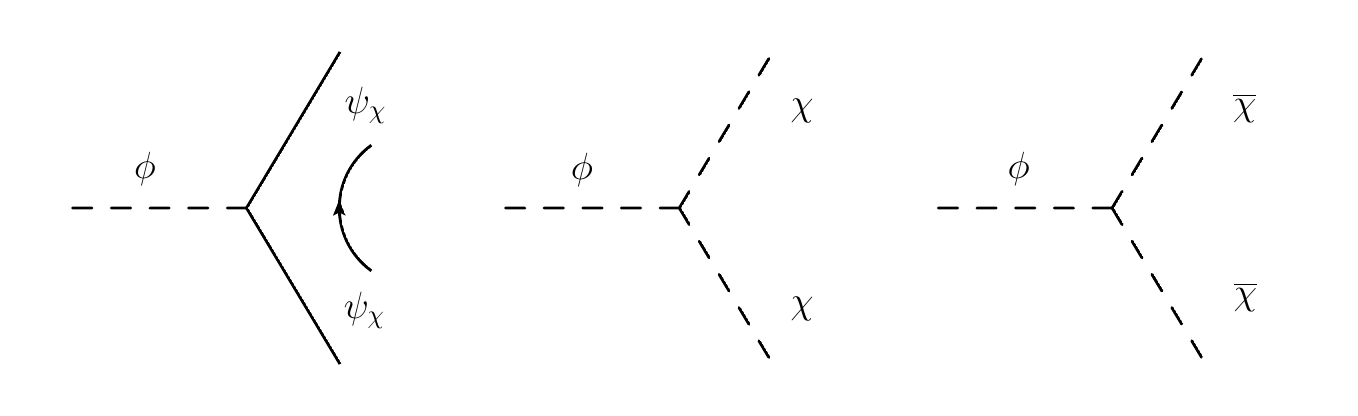} 
\caption{The three separate decay processes of the inflaton $\phi$ into right-handed neutrinos $\psi_\chi$ and right-handed sneutrinos $\chi$, $\overline{\chi}$ which contribute to the total decay rate $\Gamma_\phi$.}
\label{fig:last}
\end{figure}

The decay rate $\Gamma$ for an initial particle $i$ in the rest frame decaying into two final particles $f$ with identical mass $m_f$ has the following form: \begin{equation}\label{A1}
\begin{aligned}
\Gamma_{i \rightarrow ff}= \frac{1}{16 \pi m_i} \rvert \mathcal{M} \rvert^2 \sqrt{1-\frac{4m_f^2}{m_i^2}},
\end{aligned}
\end{equation}
where $m_i$ is the mass of the initial particle and $\mathcal{M}$ denotes the matrix element for the process. The calculation of the total decay rate implies that all separate decay channels need to be summed. In case of the inflaton which decays into right-handed (s)neutrinos this needs to be done for the processes $\phi\rightarrow \chi\, \chi$, $\phi \rightarrow \overline{\chi}\,\overline{\chi}$ and $\phi \rightarrow \psi_\chi\, \psi_\chi$ (cf.\ figure \ref{fig:last}). In the following we deduce from the superpotential (\ref{B1}) the coupling constants responsible for the inflaton decay and derive the total decay rate $\Gamma_\phi$ for general $n$. A similar derivation of the inflaton decay rate for the specific case $n=2$ can be found in \cite{Antusch:2014qqa}.

\subsection*{Couplings and matrix elements}

The inflaton-sneutrino coupling which allows the inflaton $\phi$ to decay can be obtained from the third term of the scalar potential eq.\ (\ref{B2}), where the inflaton is expanded around the vacuum expectation value $\phi = v + \delta \phi$ with small perturbations $\delta \phi$:
\begin{equation}\label{A2}
\begin{aligned}
V(\delta\phi,\chi)&=\frac{1}{2}2^{2-n} \lambda^2v^{2n}\chi^2+2^{2-n}n \lambda^2v^{2n-1}\chi^2\delta\phi+...\:.
\end{aligned}
\end{equation}
The second term gives the relevant coupling constant for the inflaton decay into sneutrinos (whereas from the first term the sneutrino mass emerges, which is identical to the one in eq.\ (\ref{M1-2})). Additional terms, such as higher order terms in $\phi$ or quartic terms in $\chi$, are denoted by '...'. The coupling which covers the inflaton to neutrino decay $\phi \rightarrow \psi_\chi \psi_\chi$ can be extracted from
\begin{equation}\label{A3}
\begin{aligned}
-\frac{1}{2}\biggr(W^{\chi\chi}\psi_\chi\psi_\chi+h.c.\biggr)=-\biggr(\frac{\lambda}{2^\frac{n}{2}} \phi^n\psi_\chi \psi_\chi+ h.c.\biggr),
\end{aligned}
\end{equation}
where  $W^{\chi\chi}$ means that the derivative is taken twice of the superpotential $W$ (given in eq.\ (\ref{B1})) with respect to the field $\chi$. The remaining superfields, i.e.\ the inflaton superfield in the present case is exchanged by its scalar component. Similar to the scalar potential, the inflaton is expanded around its minimum $\phi=v+\delta \phi$ leading to following interaction terms:
\begin{equation}\label{A4}
-2^{-\frac{n}{2}} \lambda(v^n+n v^{n-1}\delta\phi+...+\delta\phi^n)\psi_\chi \psi_\chi+ h.c.,
\end{equation}
where the second term denotes the decay channel for the inflaton into two right-handed neutrinos. By substituting the vacuum expectation value $v$ for the (s)neutrino mass $m_\chi$ (\ref{M1-2}), the following tree-level matrix elements are obtained:

\begin{eqnarray}\label{}
i\cal{M}_{\phi \, \rightarrow \chi \chi/\bar \chi  \bar \chi} &=& -i n(2^\frac{2+n}{2}\lambda m_\chi^{2n-1})^\frac{1}{n},\\
i{\cal M_{\phi \, \rightarrow \psi_\chi \psi_\chi}} &=& -i n(2^{\frac{2-3n}{2}}\lambda m_\chi^{n-1})^\frac{1}{n}\,\bar{u}(p_1,s_1)v(p_2,s_2).
\end{eqnarray}

\subsection*{The inflaton decay rate for general $n$}

The decay rate for the inflaton $\Gamma_\phi$ is obtained by summing over all squared matrix elements, which are evaluated in the centre of mass frame with $p_1=(m_\phi/2, \vec{p}_1)$ and $p_2=(m_\phi/2,-\vec{p}_1)$,
\begin{equation}\label{A5}
\Gamma_{\phi}=\frac{1}{4\pi}\biggr(\biggr(\frac{2 \lambda}{m_\chi}\biggr)^{1/n}\frac{n \, m_\chi}{4}\biggr)^2 m_\phi \biggr(\,1+12\frac{m^2_\chi}{m^2_\phi}\,\biggr)\sqrt{1-4\frac{m^2_\chi}{m^2_\phi}},
\end{equation}
where the mass of the initial particle has been identified with $m_\phi$ and the outgoing one with the (s)neutrino mass $m_\chi$. For specific cases it turns out to be useful to trade the coupling constant $\lambda$ for the right-handed (s)neutrino mass $m_\chi$ and the vacuum expectation value $v$ (see eq.\ (\ref{M1-2})):
\begin{equation}\label{A6}
\Gamma_{\phi}=\frac{n^2}{32\pi}\frac{m_\chi^2 m_\phi}{v^2}\biggr(\,1+12\frac{m^2_\chi}{m^2_\phi}\,\biggr)\sqrt{1-4\frac{m^2_\chi}{m^2_\phi}}.
\end{equation}

\begin{figure}[p]
\centering
\subfigure{%
\includegraphics[height=6.7cm]{figures/BAU_p4_preh_2.pdf}}%
\hspace{0pt}
\subfigure{%
\includegraphics[height=6.7cm]{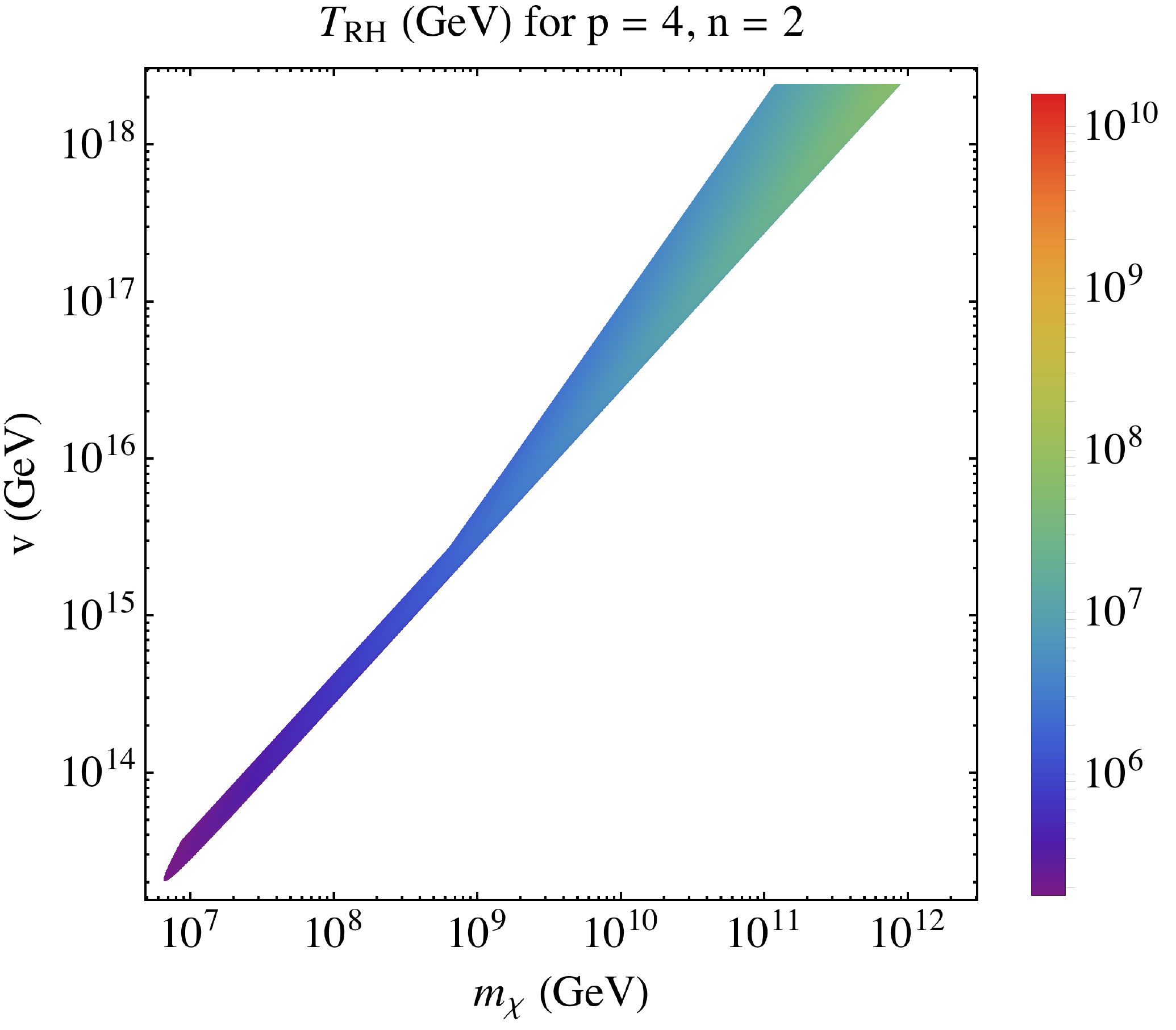}}\\
\subfigure{%
\includegraphics[height=6.7cm]{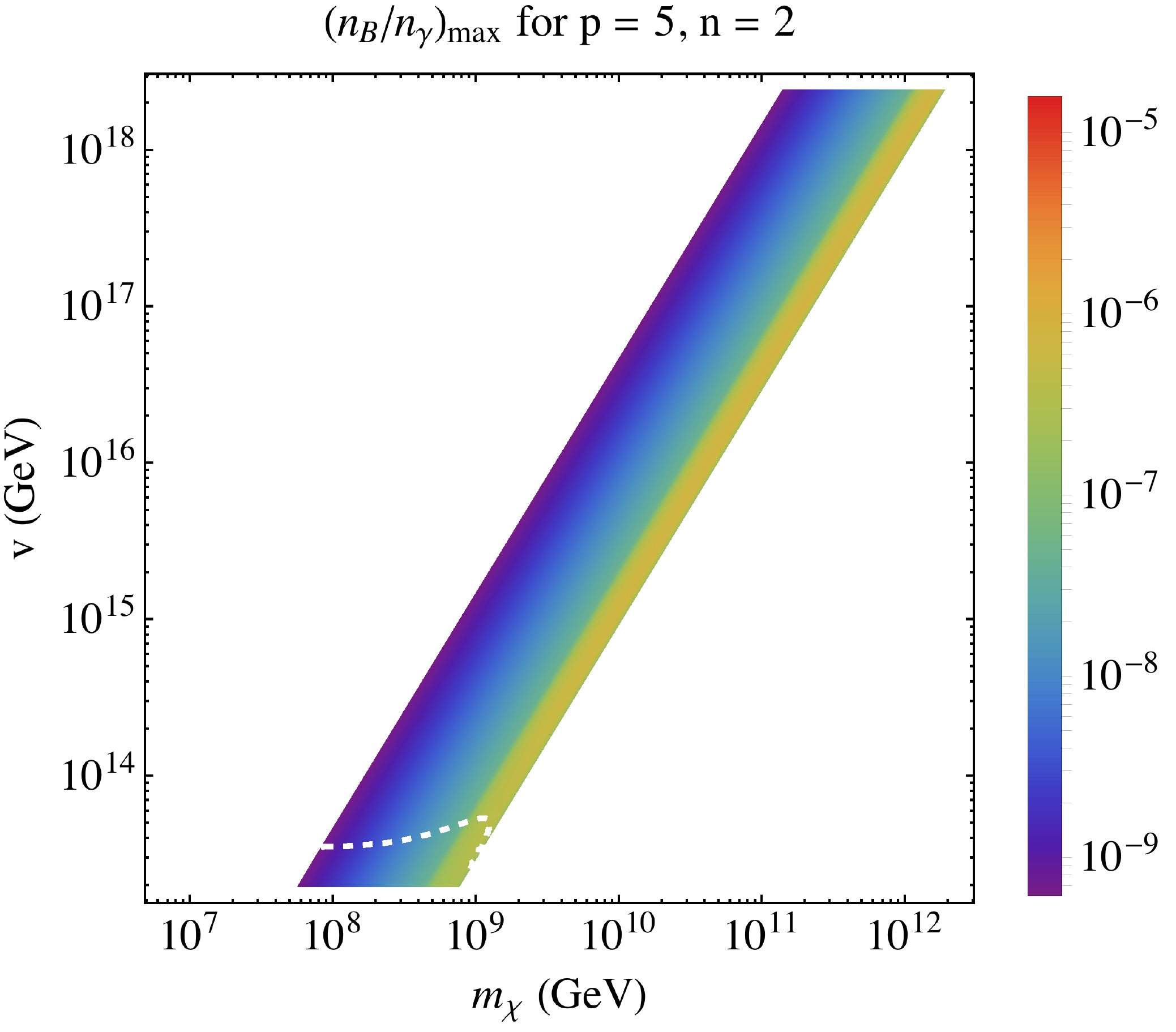}}%
\hspace{0pt}
\subfigure{%
\includegraphics[height=6.7cm]{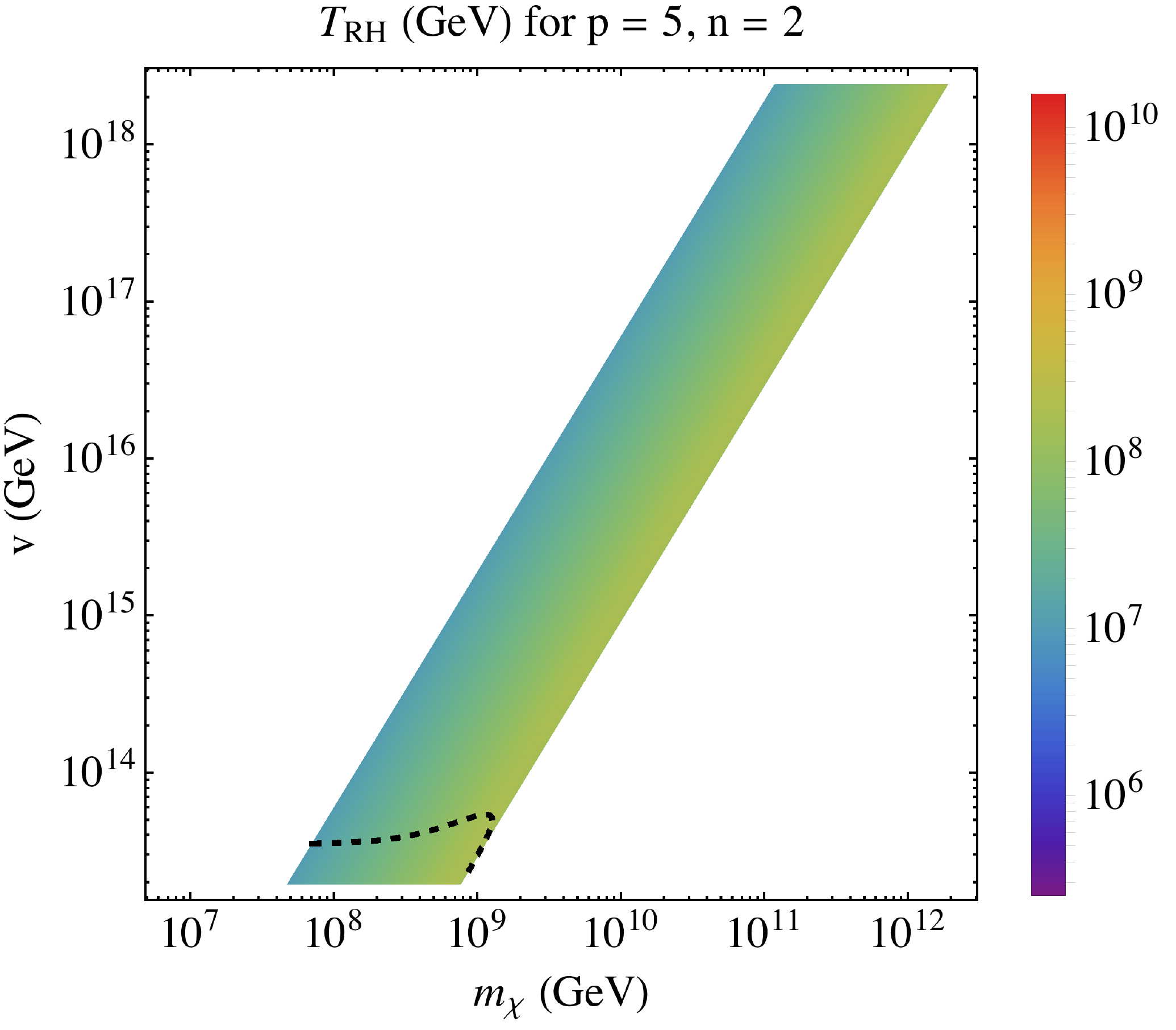}}\\
\subfigure{%
\includegraphics[height=6.7cm]{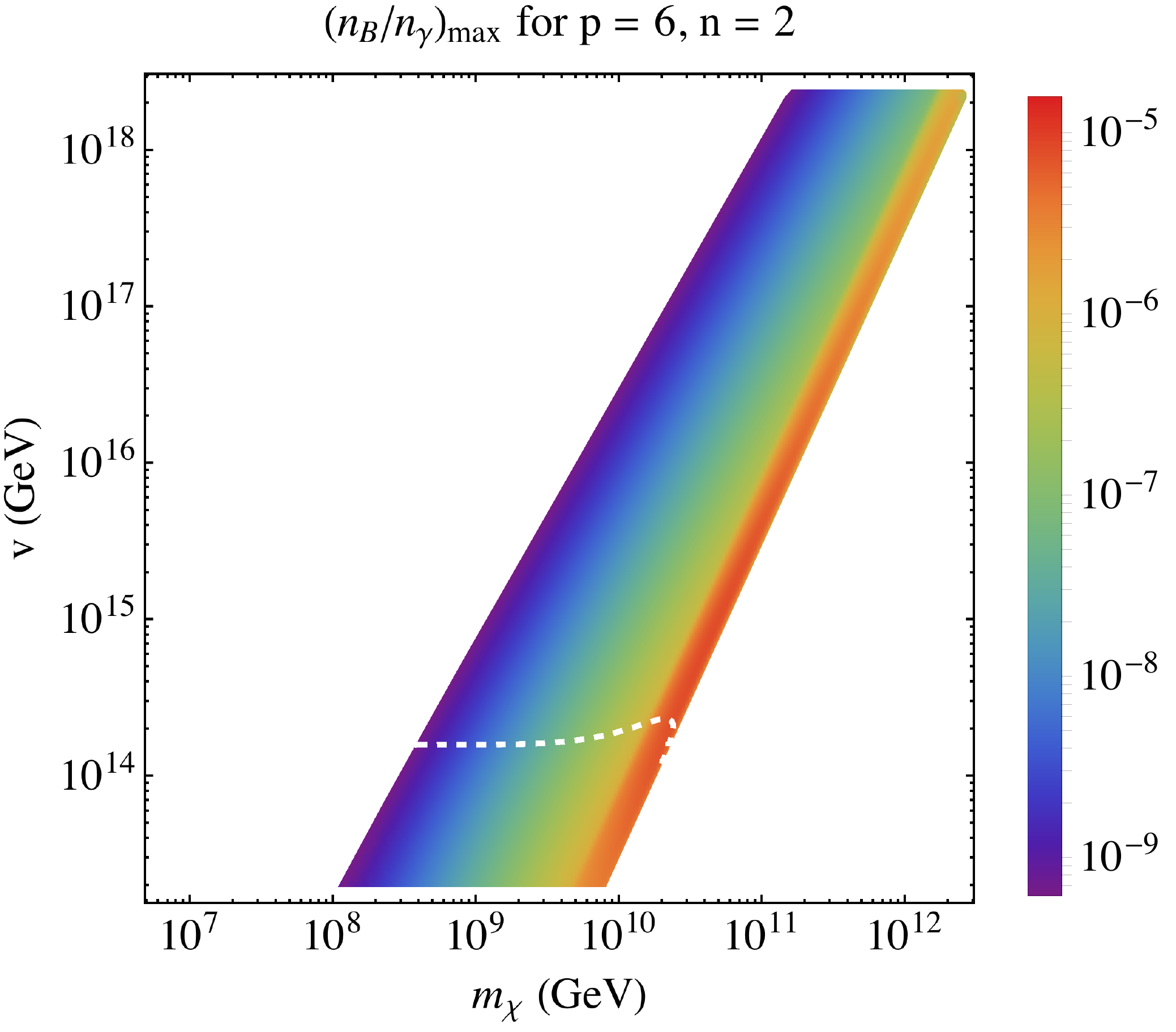}}%
\hspace{0pt}
\subfigure{%
\includegraphics[height=6.7cm]{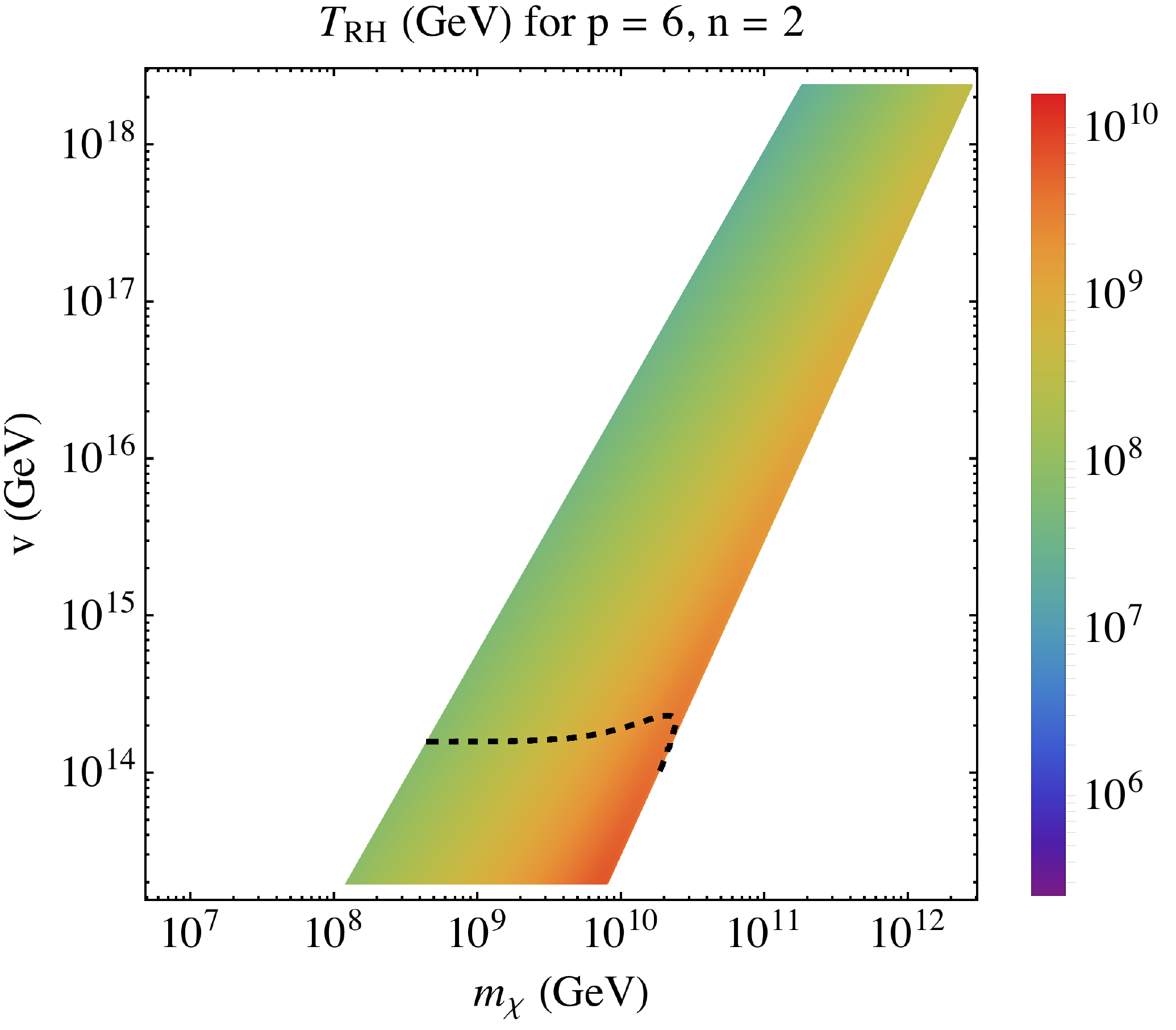}}%
\caption{Maximally produced baryon asymmetry and reheat temperature depicted as functions of right-handed (s)neutrino mass $m_\chi$ and vacuum expectation value $v$ for the cases $n=2$ and $p=4, 5, 6$. The visible areas are restricted by $m_\phi<2\,m_\chi$ on the right and $n_B/n_\gamma < 6.19 \times 10^{-10}$ on the left. The cases $p=5, 6$ contain a white/black dashed line indicating $m_\chi=10\times T_\mathrm{RH}$ where above washout effects can be neglected and leptogenesis is regarded as fully non-thermal.}
\label{fig:nB_Trh}
\end{figure}


\begin{thebibliography}{jcap}


\bibitem{inf}
A.~H. Guth,
 \textit{The inflationary universe: A possible solution to the horizon
  and flatness problems},
Phys. Rev. \textbf{D23} (1981), 347--356; 
A.~D. Linde,
 \textit{A new inflationary universe scenario: A possible solution of
  the horizon, flatness, homogeneity, isotropy and primordial monopole
  problems},
Phys. Lett. \textbf{B108} (1982), 389--393; 
A.~Albrecht and P.~J. Steinhardt,
 \textit{Cosmology for grand unified theories
  with radiatively induced symmetry breaking},
Phys. Rev. Lett. \textbf{48}
(1982), 1220--1223; 
For a review containing an extensive list of references,
see e.g.: D.~H.~Lyth and A.~Riotto,
  \textit{Particle physics models of inflation and the cosmological density
perturbation}, 
Phys.\ Rept.\  {\bf 314} (1999) 1 [hep-ph/9807278].


\bibitem{Senoguz:2004ky}
  V.~N.~Senoguz and Q.~Shafi,
  \textit{New inflation, preinflation, and leptogenesis},
  Phys.\ Lett.\ B {\bf 596} (2004) 8
  [hep-ph/0403294].

\bibitem{Nakayama:2011ri}
  K.~Nakayama and F.~Takahashi,
  \textit{Low-scale Supersymmetry from Inflation},
  JCAP {\bf 1110} (2011) 033
  [arXiv:1108.0070 [hep-ph]].
  
\bibitem{Antusch:2014qqa}
	S.~Antusch, D.~Nolde and S.~Orani,
	\textit{Hilltop inflation with preinflation from coupling to matter fields},
	JCAP {\bf 1405} (2014) 034
	[arXiv:1402.5328 [hep-ph]].
	
	 \bibitem{Boucenna:2014uma}
  S.~M.~Boucenna, S.~Morisi, Q.~Shafi and J.~W.~F.~Valle,
  \textit{Inflation and majoron dark matter in the seesaw mechanism},
  Phys.\ Rev.\ D {\bf 90} (2014) no.5,  055023
  [arXiv:1404.3198 [hep-ph]].

\bibitem{King:2017nbl}
  S.~F.~King and P.~O.~Ludl,
  \textit{Supersymmetric Majoron Inflation},
  JHEP {\bf 1703} (2017) 174
  [arXiv:1701.04794 [hep-ph]].
 
 
  
 
 	\bibitem{leptogenesis}
  M.~Fukugita and T.~Yanagida,
  \textit{Baryogenesis Without Grand Unification},
  Phys.\ Lett.\ B {\bf 174} (1986) 45;
  V.~A.~Kuzmin, V.~A.~Rubakov and M.~E.~Shaposhnikov,
  \textit{On the Anomalous Electroweak Baryon Number Nonconservation in the Early Universe},
  Phys.\ Lett.\ B {\bf 155} (1985) 36;
  M.~A.~Luty,
  \textit{Baryogenesis via leptogenesis},
  Phys.\ Rev.\ D {\bf 45} (1992) 455;
  T.~Asaka, K.~Hamaguchi, M.~Kawasaki and T.~Yanagida,
  \textit{Leptogenesis in inflaton decay},
  Phys.\ Lett.\ B {\bf 464} (1999) 12
  [hep-ph/9906366].

 
\bibitem{Antusch:2015vna}	
	S.~Antusch, F.~Cefala, D.~Nolde and S.~Orani, 
	\textit{Parametric resonance after hilltop inflation caused by an inhomogeneous inflaton field}, 
	JCAP {\bf 1602} (2016) no.02,  044 
	[arXiv:1510.04856 [hep-ph]].

 
 
\bibitem{Nolde:2013bha}
  	D.~Nolde,
  	\textit{Effects of the imaginary inflaton component in supergravity new inflation},
  	JCAP {\bf 1311} (2013) 028
  	[arXiv:1310.0820 [hep-ph]].


\bibitem{Antusch:2008pn}
	S.~Antusch, M.~Bastero-Gil, K.~Dutta, S.~F.~King and P.~M.~Kostka,
	\textit{Solving the eta-Problem in Hybrid Inflation with Heisenberg Symmetry and Stabilized Modulus},	
	JCAP {\bf 0901} (2009) 040	
	[arXiv:0808.2425 [hep-ph]].
  
  
  	
	
\bibitem{Antusch:2013eca}
  S.~Antusch and F.~Cefalà,
  \textit{SUGRA New Inflation with Heisenberg Symmetry},
  JCAP {\bf 1310} (2013) 055
  [arXiv:1306.6825 [hep-ph]].



\bibitem{Antusch:2015nla}
  S.~Antusch, D.~Nolde and S.~Orani,
  \textit{Hill crossing during preheating after hilltop inflation},
  JCAP {\bf 1506} (2015) no.06,  009
  [arXiv:1503.06075 [hep-ph]].
  
  \bibitem{Brax:2010ai}
  P.~Brax, J.~F.~Dufaux and S.~Mariadassou,
  \textit{Preheating after Small-Field Inflation},
  Phys.\ Rev.\ D {\bf 83} (2011) 103510
  [arXiv:1012.4656 [hep-th]].

\bibitem{Kofman:1994rk}
  L.~Kofman, A.~D.~Linde and A.~A.~Starobinsky,
  \textit{Reheating after inflation},
  Phys.\ Rev.\ Lett.\  {\bf 73} (1994) 3195
  [hep-th/9405187].
  
  \bibitem{Kofman:1997yn}
  L.~Kofman, A.~D.~Linde and A.~A.~Starobinsky,
  \textit{Towards the theory of reheating after inflation},
  [hep-ph/9704452].

\bibitem{Antusch:2010mv}
  S.~Antusch, J.~P.~Baumann, V.~F.~Domcke and P.~M.~Kostka,
  \textit{Sneutrino Hybrid Inflation and Nonthermal Leptogenesis},
  JCAP {\bf 1010} (2010) 006
  [arXiv:1007.0708 [hep-ph]].

\bibitem{Davidson:2008bu}
  S.~Davidson, E.~Nardi and Y.~Nir,
  \textit{Leptogenesis},
  Phys.\ Rept.\  {\bf 466} (2008) 105
  [arXiv:0802.2962 [hep-ph]].
  
  
\bibitem{Khlopov:1984pf}
  M.~Y.~Khlopov and A.~D.~Linde,
  \textit{Is It Easy to Save the Gravitino?},
  Phys.\ Lett.\  {\bf 138B} (1984) 265.
  
\bibitem{Ellis:1984eq}
  J.~R.~Ellis, J.~E.~Kim and D.~V.~Nanopoulos,
  \textit{Cosmological Gravitino Regeneration and Decay},
  Phys.\ Lett.\  {\bf 145B} (1984) 181.

\bibitem{Moroi:1993mb}
  T.~Moroi, H.~Murayama and M.~Yamaguchi,
  \textit{Cosmological constraints on the light stable gravitino},
  Phys.\ Lett.\ B {\bf 303} (1993) 289.
  
  \bibitem{Hook:2018sai}
 A.~Hook, R.~McGehee and H.~Murayama,
  \textit{Cosmologically Viable Low-energy Supersymmetry Breaking},
  arXiv:1801.10160 [hep-ph].
  
\bibitem{Kawasaki:2004yh}
  M.~Kawasaki, K.~Kohri and T.~Moroi,
  \textit{Hadronic decay of late - decaying particles and Big-Bang Nucleosynthesis},
  Phys.\ Lett.\ B {\bf 625} (2005) 7
  [astro-ph/0402490].
  
\bibitem{Covi:1996wh}
  L.~Covi, E.~Roulet and F.~Vissani,
  \textit{CP violating decays in leptogenesis scenarios},
  Phys.\ Lett.\ B {\bf 384} (1996) 169
  [hep-ph/9605319].

\bibitem{Davidson:2002qv}
  S.~Davidson and A.~Ibarra,
  \textit{A Lower bound on the right-handed neutrino mass from leptogenesis},
  Phys.\ Lett.\ B {\bf 535} (2002) 25
  [hep-ph/0202239].
  
\bibitem{Hamaguchi:2001gw}
  K.~Hamaguchi, H.~Murayama and T.~Yanagida,
  \textit{Leptogenesis from N dominated early universe},
  Phys.\ Rev.\ D {\bf 65} (2002) 043512
  [hep-ph/0109030].

  



\end{thebibliography}
\end{document}